\documentclass[pdflatex,sn-mathphys-num]{sn-jnl}

\usepackage{graphicx}%
\usepackage{multirow}%
\usepackage{amsmath,amssymb,amsfonts}%
\usepackage{amsthm}%
\usepackage{mathrsfs}%
\usepackage[title]{appendix}%
\usepackage{xcolor}%
\usepackage{textcomp}%
\usepackage{manyfoot}%
\usepackage{booktabs}%
\usepackage{algorithm}%
\usepackage{algorithmicx}%
\usepackage{algpseudocode}%
\usepackage{listings}%
\usepackage[acronym]{glossaries}%
\usepackage{csquotes}%
\makeglossaries

\usepackage{tikz}
\usetikzlibrary{shapes.geometric, arrows}
\usetikzlibrary{shapes.geometric, arrows.meta, positioning}
\tikzstyle{box} = [rectangle, rounded corners, minimum width=1cm, minimum height=1.5cm, text centered, draw=black, fill=white, line width=0.25mm]
\tikzstyle{boxhead} = [rectangle, rounded corners, minimum height=1cm, text centered, draw=black, fill=white, line width=0.25mm]
\tikzstyle{graybox} = [box, fill=gray!20]
\tikzstyle{arrow} = [thick,->,>=stealth]

\usepackage{array}
\usepackage{longtable}
\usepackage{threeparttablex} 

\lstdefinelanguage{json}{
  basicstyle=\ttfamily,
  numbers=left,
  numberstyle=\tiny,
  stepnumber=1,
  numbersep=5pt,
  showstringspaces=false,
  breaklines=true,
  frame=single,
  backgroundcolor=\color{white},
  literate=
   *{0}{{{\color{blue}0}}}{1}
    {1}{{{\color{blue}1}}}{1}
    {2}{{{\color{blue}2}}}{1}
    {3}{{{\color{blue}3}}}{1}
    {4}{{{\color{blue}4}}}{1}
    {5}{{{\color{blue}5}}}{1}
    {6}{{{\color{blue}6}}}{1}
    {7}{{{\color{blue}7}}}{1}
    {8}{{{\color{blue}8}}}{1}
    {9}{{{\color{blue}9}}}{1}
    {:}{{{\color{red}:}}}{1}
    {,}{{{\color{red},}}}{1}
    {"}{{{\color{brown}"}}}{1}
    {[}{{{\color{magenta}[}}}{1}
    {]}{{{\color{magenta}]}}}{1}
    {\{}{{{\color{magenta}\{}}}{1}
    {\}}{{{\color{magenta}\}}}}{1}
}

\newacronym{IM}{IM}{information model}
\newacronym{DM}{DM}{data model}
\newacronym{IDM}{IDM}{information and data model}
\newacronym{DR}{DR}{demand response}
\newacronym{UML}{UML}{Unified modeling language}
\newacronym{ER}{ER}{entity-relationship}
\newacronym{XML}{XML}{Extensible Markup Language}
\newacronym{JSON}{JSON}{JavaScript Object Notation}
\newacronym{DSR}{DSR}{design science research}
\newacronym{DSRM}{DSRM}{design science research methodology}
\newacronym{KM}{KM}{knowledge models}
\newacronym{EFOnt}{EFOnt}{Energy Flexibility Ontology}
\newacronym{EFDM}{EFDM}{energy flexibility data model}
\newacronym{AMOD}{AMOD}{agile methodology for ontology development}
\newacronym{QC}{QC}{quality characteristic}
\newacronym{IoT}{IoT}{internet of things}
\newacronym{BIM}{BIM}{building information model}
\newacronym{IT}{IT}{information technology}
\newacronym{ESP}{ESP}{energy synchronisation platform}
\newacronym{DSO}{DSO}{distribution system operator}
\newacronym{TSO}{TSO}{transmission system operator}
\newacronym{EFIM}{EFIM}{energy flexibility information model}
\newacronym{UUID}{UUID}{universally unique identifier}
\newacronym{EFIDM}{EFIDM}{energy flexibility information and data model}
\newacronym{SGAM}{SGAM}{Smart Grid Architecture Model}
\newacronym{NIST}{NIST}{National Institute of Standards and Technology}
\newacronym{CIM}{CIM}{Common Information Model}
\newacronym{EMS}{EMS}{energy management system}
\newacronym{API}{API}{application programming interface}
\newacronym{IEC}{IEC}{International Electrotechnical Commission}
\newacronym{CEN}{CEN}{European Committee for Standardization}
\newacronym{CENELEC}{CENELEC}{European Committee for Electrotechnical Standardization}
\newacronym{ETSI}{ETSI}{European Telecommunications Standards Institute}
\newacronym{EI}{EI}{Energy Interoperation}
\newacronym{USEF}{USEF}{Universal Smart Energy Framework}
\newacronym{UFTP}{UFTP}{USEF Flex Trading Protocol}
\newacronym{EMIX}{EMIX}{Energy Market Information Exchange}
\newacronym{LF}{LF}{Linux Foundation}
\newacronym{COSEM}{COSEM}{Companion Specification for Energy Metering}
\newacronym{DER}{DER}{Distributed Energy Resources}
\newacronym[longplural={Information and Communication Technologies}]{ICT}{ICT}{Information and Communication Technology}
\newacronym{TP}{TP}{test purpose}
\newacronym{ICS}{ICS}{Implementation Conformance Statement}
\newacronym{ATS}{ATS}{Abstract Test Suite}
\newacronym{PIM}{PIM}{platform independent model}
\newacronym{PSM}{PSM}{platform specific model}
\newacronym{CIM_MDA}{CIM}{computation independent model}
\newacronym{CoC}{CoC}{Code of Conduct}
\newacronym{JRC}{JRC}{Joint Research Centre}
\newacronym[shortplural={IMs \& DMs}]{IDM1}{IM \& DM}{information and data model}

\begin{document}

\title[Evaluation Method for Information and Data Models]{Enhancing Smart Grid Information Exchanges: A Three-Phase Method for Evaluating Information and Data Models during their Development Process}

\author*[1]{\fnm{Christine} \sur{van Stiphoudt}}\email{christine.vanstiphoudt(at)uni.lu}

\author[1]{\fnm{Sergio} \sur{Potenciano Menci}}\email{sergio.potenciano-menci(at)uni.lu}

\author[1]{\fnm{Gilbert} \sur{Fridgen}}\email{gilbert.fridgen(at)uni.lu}

\affil*[1]{\orgdiv{SnT - Interdisciplinary Centre for Security, Reliability and Trust}, \orgname{University of Luxembourg}, \orgaddress{\street{29, Avenue J.F Kennedy}, \city{Luxembourg}, \postcode{L-1855}, \country{Luxembourg}}}


\abstract{

The ongoing process of smart grid digitalisation is increasing the volume of automated information exchange across distributed energy systems. 
This has driven the development of new \acrlongpl{IDM1} when existing models fail to offer an optimal description of the requisite information due to be exchanged. 
To prevent potential operational disruption – i.e. in the provision of flexibility – caused by flaws in these newly designed models, it is essential to conduct evaluations during the development process before these models are deployed.

Current practices differ across domains. Beyond smart grid applications, information models are evaluated through explicit reviews using \acrlongpl{QC}. Within smart grid contexts, evaluation focuses on data models and implicit system-level conformance and interoperability testing. However, no existing approach combines these explicit and implicit evaluation methods for both \acrlongpl{IDM1} during their development. This limits early fault detection and increases potential model correction costs. 

To address this gap, we propose a three-phase evaluation method based on design science research. Our method integrates explicit and implicit approaches, applies them to \acrlongpl{IDM1} and is adaptable to various design stages. We also introduce a set of \acrlongpl{QC} to support explicit model evaluation. Overall, our contribution enhances the reliability and interoperability of smart grid information exchange.

}


\maketitle

\section{Introduction}\label{sec:intro}

The ongoing process of smart grid digitalisation is increasing the volume of automated information exchange between distributed energy systems~\cite{TUBALLA2016710}. It has led to the development of new \glspl{IDM1} where existing models cannot sufficiently describe the information to be exchanged~\cite{7947600}. 
For instance, in the context of energy flexibility and \gls{DR}, well-known examples of such new \glspl{IDM1} include openADR~\cite{2015_openADRAlliance_OpenADR20Profile} and EEBus SPINE~\cite{2023_EEBusInitiativee.V._EEBusSPINETechnical}. These have been developed to support the integration and control of distributed energy resources for the provision of flexibility. 

However, design flaws in these emerging models can lead to operational disruptions in the provision of flexibility. Therefore, evaluating \glspl{IDM1} during their development phase is essential to ensure reliability prior to operational deployment. 
Yet, we observe that existing model descriptions frequently lack detail on how such evaluations were conducted~\citep{2015_openADRAlliance_OpenADR20Profile, 2023_EEBusInitiativee.V._EEBusSPINETechnical}.
As a result, model developers working on the integration of distributed energy resources and flexibility have limited guidance on how to evaluate their models during the development process. This challenge is particularly critical for industrial flexibility contexts given the high potential for flexibility in industrial systems~\cite{2019_Schott_GenericDataModel}.

According to the existing literature (see Section~\ref{sec:ModelQualEval}), evaluation approaches typically rely on explicit validation of \textit{conceptual model quality}. These approaches focus on making direct assessments of \glspl{IM} – i.e. the conceptual specification of data structures – by defining \glspl{QC} and evaluation frameworks~\citep{2026_vanStiphoudt_ExAnteEvaluationApproaches}. However, existing explicit approaches rarely provide practical step-by-step guidance. 
In contrast, evaluation practices in smart grid contexts tend to be implicit, testing conformance and interoperability of systems~\cite{2018_Papaioannou_SmartGridInteroperability}. Such approaches focus on evaluating \textit{system functionality} in interactions by testing \gls{DM} instances – i.e. implementation-level specifications of the data format – and thus assessing models indirectly as a by-product~\citep{2012_ETSI_MethodsTestingSpecification}. Implicit evaluations are applied post-deployment, and therefore offer limited support for evaluating models during their development process.

Broader literature on standard evaluation (see Section~\ref{sec:ModelQualEval}), which is relevant for \glspl{IDM1}, suggests that combining explicit and implicit validation can enhance model evaluation~\cite{2012_ETSI_MethodsTestingSpecification}. However, no existing approach integrates both in a structured and actionable manner suitable for use during the model development process in smart grid contexts. 

This gap motivates this manuscript's central research question:
\textit{how can \glspl{IDM1} used in smart grid interactions be evaluated during their development process to ensure both conceptual model quality and system functionality?} 
Our contributions towards answering this question are as follows:

\begin{itemize}
    \item An \textit{overview of existing approaches} related to \gls{IDM1} evaluation, considering both explicit and implicit model evaluation perspectives.
    
    \item A \textit{three-phase evaluation method} designed using \gls{DSR}, which combines explicit and implicit model validation. Our method includes model scope definition (Phase 1), explicit model evaluation through conceptual assessment of  \glspl{IDM1} quality (Phase 2), 
    and implicit model evaluation through testing of system functionality (Phase 3).
    
    \item A \textit{set of \glspl{QC}} to support  evaluation in Phase 2. We consolidate sets of \glspl{QC} identified in the reviewed literature and those identified in an observation-based approach~\cite{2005_Moody_TheoreticalPracticalIssues}.
    We also analyse which \glspl{QC} – applicable for \glspl{IM} – are also relevant for the evaluation of \glspl{DM}.
    
\end{itemize}

We organise the remainder of our manuscript as follows.
In Section~\ref{sec:background}, we provide the conceptual background of information exchange in smart grids and related model terminology.
In Section~\ref{sec:ModelQualEval}, we review existing literature in the context of \gls{IDM1} evaluation, highlighting current practices and their limitations.
In Section~\ref{sec:researchApproach}, we describe our research approach.
In Section~\ref{sec:evalApproach}, we propose our three-phase evaluation method for \glspl{IDM1} used in smart grid interactions during the development process.   
We demonstrate the applicability of our method by evaluating the \gls{EFIM} and \gls{EFDM} during their development processes.
In Section~\ref{sec:QC}, we introduce a set of \glspl{QC} designed to complement our evaluation method.
In Section~\ref{sec:eval}, we describe how we evaluate our designed evaluation method.
In Section~\ref{sec:discussion}, we reflect on the main implications of our research, discuss its limitations, and suggest potential avenues for future research.
In Section~\ref{sec:conclusion}, we summarise our main findings.

\section{Background} \label{sec:background}

Given that disciplines employ specific terminology~\cite{2022_Cabot_ModelingShouldBe}, establishing a clear terminology is paramount when discussing modelling across multiple domains.
In our manuscript, we adopt the interdisciplinary perspective of energy informatics, combining approaches used in energy engineering, information systems engineering, as well as systems and software engineering~\citep{2012_Watson_EnergyInformaticsInitial, 2015_USLAR_EnergyInformaticsDefinition}. Consequently, in Section~\ref{subsec:background_infExch}, we outline the scope of the models to which our evaluation method can be applied, and in Section~\ref{subsec:modelDef} we define the terms \gls{IDM1}, which we use throughout our manuscript.

\subsection{Information exchange in the smart grid}
\label{subsec:background_infExch}

We use the \gls{SGAM} framework~\cite{2014_CEN_SGAMUserManual} to define the scope of our work in the energy informatics domain.  
The \gls{SGAM} framework combines the smart grid plane with the concept of interoperability layers.  
The smart grid plane consists of five \textit{domains} that describe the energy conversion chain (i.e. generation, transmission, distribution, \gls{DER}, customer premises), and the six \textit{zones} that describe hierarchical levels of information management along the electrical processes~\cite{2014_CEN_SGAMUserManual} (i.e. process, field, station, operation, enterprise, market). The smart grid plane is part of each of the five interoperability layers (i.e. business, function, information, communication, and component), which need to be considered when aiming for interoperable end-to-end system interaction~\cite{2014_CEN_SGAMUserManual}.

\begin{figure}[ht!]
  \centering
  \includegraphics[width=0.6\textwidth]{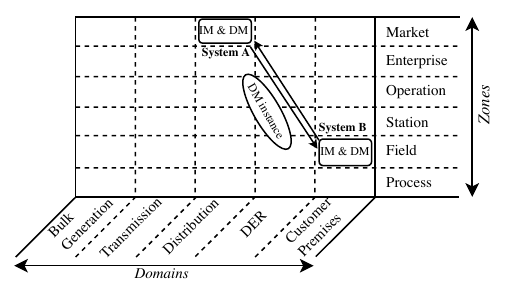}
  \caption{Simplified visualisation of the information exchange between two systems in the \gls{SGAM} information layer.}
  \label{fig:problemStatement}
\end{figure}

As our focus is the evaluation of \acrshortpl{IM} and \acrshortpl{DM} used for automated information exchange in smart grids, we address specifically the information interoperability layer. 
While communication protocols and hardware – as part of the communication and component interoperability layers – are essential for enabling communication, they fall outside the scope of this manuscript. 
We illustrate the use of both model types in system interactions for information exchange in Fig.~\ref{fig:problemStatement}.
Although information systems have access to both \gls{IDM1} specifications for automated information processing, they only exchange \acrshort{DM} instances.

\subsection{Model type definitions}
\label{subsec:modelDef}

In this manuscript, we focus on static models as opposed to models that specify behaviour. We understand static models as those which specify descriptive characteristics of a system~\cite{1995_Levitin_QualityDimensionsConceptual}. 
In Table~\ref{tab:modelDef}, we list a selection of model types and their definitions that we encountered during our research. We did not seek to compile an exhaustive list, but rather offer an overview of domain-specific perspectives and explain the reasoning behind the terminology used in this manuscript.

\begin{table}[!ht]
\centering
\caption{Model type definitions.}
\label{tab:modelDef}
\small
\begin{tabular}{p{0.15\textwidth}p{0.35\textwidth}p{0.04\textwidth}p{0.15\textwidth}p{0.15\textwidth}}
\toprule
Model type & Definition & Year & Domain & Reference \\ 
\midrule

    \multirow{2}{*}{\parbox[t]{2cm}{\centering Conceptual view}}
         & \enquote{A collection of objects representing the entities, properties and relationships of interest in the enterprise} & 1978 & Database Management & \citet{1978_Tsichritzis_ANSIX3SPARC}\\
        & \enquote{A model about the part of the real world captured in the data}; \enquote{a static model of reality} & 1994 & Database Management & \citet{1995_Levitin_QualityDimensionsConceptual}\\

\midrule

    \multirow{1}{*}{\parbox[t]{2cm}{\centering Conceptual model}}
        & \enquote{Any collection of specification statements relevant to some problem} & 1994 & Requirements Engineering & \citet{1994_Lindland_UnderstandingQualityConceptual} \\
        &  \enquote{A model of a domain made in a formal or semi-formal language with a limited vocabulary.} & 2012 & Information systems&\citet{2012_Krogstie_ModelbasedDevelopmentEvolution} \\
\midrule 

    \multirow{1}{*}{\parbox[t]{2cm}{\centering Information model}}
         & \enquote{A representation of concepts, relationships, constraints, rules, and operations to specify data semantics for a chosen domain of discourse} & 1999  & Information exchange in manufacturing &\citet{1999_Lee_InformationModelingDesign} \\
        & \enquote{[...] model[s] managed objects at a conceptual level, independent of any specific implementations or protocols used to transport the data.} 
        & 2003 
        & Network management 
        &\citet{2003_Pras_DifferenceInformationModels} (RFC 3444) \\
         
\midrule 
        
    \multirow{1}{*}{\parbox[t]{2cm}{\centering Data model}}
        & \enquote{[...] define[s] managed objects at a lower level of abstraction. [It] include[s] implementation- and protocol-specific details [...].}
        & 2003  & Network management &\citet{2003_Pras_DifferenceInformationModels} (RFC 3444) \\
\midrule 
    \parbox[t]{2cm}{\centering Platform independent model (PIM)}
    & \enquote{A model that is independent of [...] a platform}
    &  2014 & Model Driven Architecture&\citet{2014_ObjectManagementGroup_ModelDrivenArchitecture}\\
\midrule 
    \parbox[t]{2cm}{\centering Platform specific model (PSM) }       
    & \enquote{A model of a system [that] is defined in terms of a specific platform} & 2014 & Model Driven Architecture  &\citet{2014_ObjectManagementGroup_ModelDrivenArchitecture}\\

\bottomrule
\end{tabular}%
\end{table}

A widely cited publication in the field of database management – with over 500 citations – is~\citet{1978_Tsichritzis_ANSIX3SPARC}. This work defines the term \textit{conceptual view} as a \enquote{collection of objects representing the entities, properties and relationships of interest in the enterprise}. 
\citet{1995_Levitin_QualityDimensionsConceptual} discuss the representation of reality in models and define a conceptual view as a \enquote{static model of reality}. In related fields – such as requirements engineering and information systems – the term \textit{conceptual model} is more commonly used. 
For instance,~\citet{2012_Krogstie_ModelbasedDevelopmentEvolution} defines a conceptual model as a domain model specified in a formal or semi-formal language. 
A closely related definition used in the literature is provided for \textit{\acrshortpl{IM}}~\cite{2003_Pras_DifferenceInformationModels}. 
For instance, \citet{1999_Lee_InformationModelingDesign} defines \acrshortpl{IM} as \enquote{a representation of concepts, relationships, constraints, rules and operations that specify data semantics for a chosen domain}. Similarly, in the context of network management, \citet{2003_Pras_DifferenceInformationModels} highlight the conceptual nature of these models and the independence of implementation- or protocol-specific details. Conversely, \textit{\acrshortpl{DM}} contain such implementation- or protocol-specific details, and are thus specified at a lower level of abstraction. Although the term \acrshort{DM} is frequently used in the reviewed literature, it is frequently employed without a clear definition, and is used interchangeably with the definition of conceptual models or \acrshortpl{IM}, as observed in~\citet{2003_Moody_ImprovingQualityData}.

In the context of model-driven development, the terms \textit{\gls{PIM}} and \textit{\gls{PSM}} are commonly used to distinguish between abstract, platform-independent representations and those tailored to specific implementation environments. Although the distinction between \gls{PIM} and \gls{PSM} may initially appear to be similar to that between \glspl{IDM1}, neither term can be assigned exclusively to the information layer in the \gls{SGAM}. According to the classification by~\citet{2019_Uslar_ApplyingSmartGrid}, \gls{PIM} spans the information, communication, and component layers. It describes the architecture without providing detailed technical specifications about the components. In contrast, \gls{PSM} – which includes platform-specific implementation details – is typically located within the component layer~\citep{2019_Uslar_ApplyingSmartGrid}.

We adopt the terminology of \glspl{IDM1} as defined by~\citet{2003_Pras_DifferenceInformationModels} given the practical importance of distinguishing between conceptual and implementation-specific models. This distinction facilitates structured model development and systematic model evaluation. 
In addition, the term \acrshort{IM} is more prevalent in the context of the smart grid domain than the term conceptual model. This prevalence is reflected in the names assigned to widely used standards, such as \gls{CIM} or \gls{EI}. 
In this manuscript we use the following definitions for \glspl{IDM1}.

\acrshortpl{IM} define the structure of data~\cite{2020_Stuckenholz_BasiswissenEnergieinformatikLehr}. They model objects at a conceptual level that is implementation- and protocol-agnostic~\cite{2003_Pras_DifferenceInformationModels}, as well as being easier for humans to understand. 
The level of abstraction defined in an \gls{IM} is determined by the modelling needs of its designers~\cite{2003_Pras_DifferenceInformationModels}.
\acrshortpl{IM} can be described informally using natural languages (e.g. English) or using formal or semi-formal structured languages (e.g. \gls{UML}). 

\acrshortpl{DM} define the format of data facilitating automated exchange and processing~\cite{2020_Stuckenholz_BasiswissenEnergieinformatikLehr}. 
\acrshortpl{DM} consider a lower level of abstraction and include implementation-specific information~\cite{2003_Pras_DifferenceInformationModels}. 
A single \acrshort{IM} can be mapped onto multiple \acrshortpl{DM}~\cite{1999_Lee_InformationModelingDesign}.
However, for reliable information and data exchange, information systems must agree on the same interpretation rules~\cite{1996_Rahimifard_MethodologyDevelopEXPRESS}, thus using the same \acrshort{DM}. \acrshortpl{DM} can be described in schema specifications using markup languages such as \gls{XML} or notations such as \gls{JSON}~\cite{2020_Stuckenholz_BasiswissenEnergieinformatikLehr}. 
We refer to the result of mapping data to a \acrshort{DM} as \textit{\acrshort{DM} instance}.

\section{State of the art in model evaluation, validation and testing} 
\label{sec:ModelQualEval}

In this section we review the literature on implicit and explicit evaluation approaches relevant to \glspl{IDM1} in the interdisciplinary field of energy informatics. While explicit evaluation approaches have previously been analysed from a software engineering perspective~\citep{2026_vanStiphoudt_ExAnteEvaluationApproaches}, we extend this overview by incorporating implicit evaluation approaches and perspectives specific to the energy and smart grid domains.
The description of our approach to searching the literature is detailed in Section~\ref{subsec:LitReview}, and we provide an overview of validation approaches for standards relevant to \glspl{IDM1} in Section~\ref{subsec:ValidationStandards}. In Section~\ref{subsec:QualityModels}, we review explicit model evaluation approaches that focus on defining and evaluating the quality of conceptual models. 
In Section~\ref{subsec:TestingModel}, we review implicit evaluation approaches that focus on testing model implementations for conformance and interoperability.
To ensure clarity of the terms evaluation and validation, we adopt the definitions from~\citet{2017_ISO/IEC/IEEE24765:2017E_ISOIECIEEE} and the \gls{ETSI}~\cite{2012_ETSI_MethodsTestingSpecification}. 

\textit{Evaluation} refers to the assessment of the extent to which criteria are met; \textit{validation} is the process of determining whether requirements are met; and \textit{testing} is a specific method used to perform validation.

\subsection{Literature search approach}
\label{subsec:LitReview}

We conduct a literature review that combines integrative and narrative literature review techniques. 
We synthesise literature on the subject of \gls{IDM1} evaluation across the many domains of the interdisciplinary field of energy informatics.
For the integrative review, we adopt the guidelines proposed by~\citet{2005_Torraco_WritingIntegrativeLiterature}, while for the narrative review, we adopt the guidelines of~\citet{2006_Green_WritingNarrativeLiterature}.
The purpose of the integrative review is to explore different perspectives on the topic, break down discipline-specific silos, and redirect research efforts~\cite{2023_Cronin_WhyHowIntegrative}.
Given the complexity of the topic, we decided to incorporate a narrative review within the integrative literature review. Narrative reviews can support integrative reviews by focusing on the synthesis of the different categories and making the integrative review more manageable~\cite{2023_Cronin_WhyHowIntegrative}. 

To enhance transparency, we use a protocol for the search and analysis process as suggested by~\citet{SNYDER2019333}. Although we describe a structured approach, we do not intend to conduct a systematic review for numerous reasons: the complexity of the topic; the resources required; and the need to define explicitly inclusion and exclusion criteria in order to drastically reduce the sample size~\cite{grant2009typology}. We apply the protocol illustrated in Fig.~\ref{fig:search_protocol}.

\begin{figure}[ht] 
    \centering
    \tikzstyle{box} = [rectangle, rounded corners, minimum height=0.75cm, text centered, draw=black, fill=white, line width=0.25mm]
    \tikzstyle{arrow} = [thick,->,>=stealth]
    
    \begin{tikzpicture}[node distance=.5cm]
    
    \small
    
    \node (topic) [box, minimum width=70mm] at (0, 0) {Stream}; 
    \node (academic) [box, minimum width=30mm,align=center] at (-1.923, -1.4) {Selected databases \\ \& search term};           
    \node (nonacademic) [box, minimum width=30mm] at (1.923, -1.4) {Selected platforms};     
    \node (initial) [box, minimum width=70mm] at (0, -2.6) {Analyse literature}; 
    \node (research) [box, minimum width=70mm] at (0, -3.8) {Search \& analyse with \textit{snowballing}}; 
    \node (final) [box, minimum width=70mm] at (0, -5.0) {Final set of literature}; 

    \draw [arrow] (topic.south) ++(-1.923, 0) -- (academic.north) node[midway, right] {Academic};;
    \draw [arrow] (topic.south) ++(1.923, 0) -- (nonacademic.north) node[midway, right] {Grey};
    \draw [arrow] (academic.south) -- ([xshift=-1.923cm]initial.north);
    \draw [arrow] (nonacademic.south) -- ([xshift=1.923cm]initial.north);
    \draw [arrow] (initial.south) -- (research.north);
    \draw [arrow] (research.south) -- (final.north);
    
    \end{tikzpicture}
    \caption{Structured search and analysis protocol for combined literature review.   
    }
    \label{fig:search_protocol}
\end{figure}  

We explore two streams, i.e. academic and grey literature. In the academic literature, we use IEEE, Scopus and ACM databases, employing combinations of the following search terms: \textit{data model/information model/conceptual model}, \textit{quality/evaluation/validation/testing}, 
\textit{interoperability/conformance/conformity}, 
\textit{smart grid}. Our focus is on peer-reviewed journal articles and conference proceedings written in English.
For grey literature, we searched Google Scholar and the websites of standardisation bodies and research projects such as CORDIS~\cite{_EuropeanCommission_CORDISEUResearch} for EU projects. We focus on standard specifications and reports.

In our analysis, we select the literature based on its relevance to the overarching topic. Where deemed appropriate by the authors, we analysed both the core components and the complete manuscript.

To enhance the effectiveness of our search and to discover additional relevant literature related to our topic, we conducted a \textit{snowballing} search, as described by~\citet{2014_Callahan_WritingLiteratureReviews}. 
By tracking the references in the articles and examining their citations, we aimed to increase the effectiveness of these searches and thus find additional literature relevant to our topic.

\subsection{Validation methods for standards specifying physical characteristics}
\label{subsec:ValidationStandards}

The process for validating standards seeks to ensure that their specified requirements fulfil the intended objectives~\cite{2012_ETSI_MethodsTestingSpecification}. Examples of standards specifying \glspl{IDM1} are \gls{CIM}~\cite{2025_entsoe_CommonInformationModel} and OpenADR 2.0b ~\cite{2019_openADRAlliance_OpenADR20bReceives}.
The standardisation body~\citet{2012_ETSI_MethodsTestingSpecification}  differentiates between the validation of standards specifying physical characteristics and standards specifying behaviour. 
Since we define \glspl{IDM1} in Section~\ref{subsec:modelDef} as static models that specify descriptive characteristics rather than dynamic behaviour, we refer in this manuscript to validation approaches applicable for standards that specify physical characteristics. 

\citet{2012_ETSI_MethodsTestingSpecification} recommends explicit validation using four-step review processes: 1) planning the reviews, 2) preparing the review, 3) reviewing the specification, and 4) processing change proposals. While these reviews provide direct quality assessments of the specifications, implicit validation approaches – where the validation of the specification is only a by-product~\citep{2012_ETSI_MethodsTestingSpecification} – are also mentioned. Such implicit validation approaches include conformance and interoperability testing, both of which are essential for ensuring interoperable system interaction in a smart grid. 
Furthermore, ~\citet{2012_ETSI_MethodsTestingSpecification} recommends the use of validation methods at different stages of the development process.

To summarise, the validation methods suggested for standards specifying physical characteristics encompass both explicit and implicit validation approaches. However, the review process for explicit validation does not specifically address the characteristics of \glspl{IDM1}.

\subsection{Explicit model evaluation approaches}
\label{subsec:QualityModels}

Existing explicit model evaluation approaches aim to assess the quality of conceptual models.
Several standards define the term \textit{quality}. ISO 9000 addresses \textit{quality management}, ISO/IEC 9126 deals with software product quality, and ISO 2510 addresses the product quality of \gls{ICT} and software products. However, even if \glspl{IDM1} can adequately be placed in the software products category, no quality standard specifically addresses their unique characteristics. \citet{2005_Moody_TheoreticalPracticalIssues} proposes the following definition for \textit{conceptual model quality} that is consistent with the ISO 9000 definition: \enquote{\textit{The totality of features and characteristics of a conceptual model that bear on its ability to satisfy stated or implied needs}}. In our manuscript, we refer to these characteristics as \glspl{QC}~\cite{1995_Levitin_QualityDimensionsConceptual}. The literature also refers to the terms: \textit{quality goals}~\cite{1994_Lindland_UnderstandingQualityConceptual},  \textit{quality factors}~\cite{2003_Moody_ImprovingQualityData} or \textit{attributes}~\cite{2009_Mehmood_DataQualityConceptual}.
In this section, we synthesise explicit model evaluation approaches that take a systems and software engineering perspective.
In particular, in Section~\ref{subsubsec:QualCrit}, we analyse approaches that define \glspl{QC} for conceptual models, which we refer to as \acrshortpl{IM} (see Section~\ref{sec:background}). 
In Section~\ref{subsubsec:EvalFrame}, we analyse approaches that focus on model evaluation using these characteristics.

\subsubsection{Quality characteristics}
\label{subsubsec:QualCrit}

In this section, we focus on literature that relates directly to the quality of conceptual models, and therefore exclude literature on data/information quality such as~\citet{2009_Batini_MethodologiesDataQuality} and~\citet{2024_Helskyaho_DefiningDataModel}. 

The analysed literature can be grouped into sets of \glspl{QC}, either as theoretical frameworks or as lists. Theoretical frameworks relate multiple \glspl{QC} to each other, while lists only enumerate them. 
For example, \citet{1994_Lindland_UnderstandingQualityConceptual} define and relate model quality using four \glspl{QC} (termed quality goals) based on linguistic theory.
The SEQUAL framework extends the number of quality goals: referred to as quality levels~\cite{2013_Krogstie_QualityConceptualData}. These levels synthesise multiple perspectives on model quality, and are based on semiotic theory, a field concerned with signs and their meanings.
Similarly, the CMQF framework builds on the work of~\citet{1994_Lindland_UnderstandingQualityConceptual} and its subsequent extensions, combining it with the ontological model of~\citet{1990_Wand_OntologicalModelInformation}. It defines twenty-four \glspl{QC}, which are mapped to four layers, with these representing the basic modelling process and the object of interest.
The three theoretical frameworks \cite{1994_Lindland_UnderstandingQualityConceptual, 2012_Nelson_ConceptualModelingQuality, 2013_Krogstie_QualityConceptualData} mention syntactic correctness or syntactic quality as one \gls{QC}, emphasising its importance.  

Unlike theoretical frameworks, lists enumerate individual \glspl{QC}, but without establishing relationships among them.  
For example, \citet{1995_Levitin_QualityDimensionsConceptual} defines fourteen \glspl{QC}, illustrating them with examples from database management.
\citet{1999_Lee_InformationModelingDesign} mentions seven \glspl{QC} based on experience of model development in the manufacturing sector.
\citet{2003_Moody_ImprovingQualityData} define eight \glspl{QC}, which they apply in field and laboratory experiments. 
While all three lists are practically orientated, our analysis reveals an overlap of nine \glspl{QC} between them. 

Given the large number of existing sets of \gls{QC}, researchers have attempted to consolidate them through categorisation~\cite {1995_Levitin_QualityDimensionsConceptual, 2009_Mohagheghi_DefinitionsApproachesModel}. 
However, the reasoning behind these categories is not always clear. To avoid excessive classification which bring limited benefits, \citet{2021_Haug_UnderstandingDifferencesData} emphasises, in the context of data quality classification, the need for justification when specific sets of \glspl{QC} are created.
Additionally, some papers have proposed quantitative metrics to offer objective measurements of \glspl{QC}. Given that practitioners have found these metrics to offer only marginal benefits for improving \acrshortpl{IM}~\cite{2003_Moody_MeasuringQualityData}, we exclude them from our analysis.

To summarise, while the sets of \glspl{QC} defined within theoretical frameworks are often abstract, those defined in lists tend to be more specific and therefore more useful for practical applications~\cite{1997_Shanks_QualityConceptualModelling}. However, their definitions are often vague, complicated, or even absent~\cite{1994_Lindland_UnderstandingQualityConceptual}, and may vary between different studies~\cite{2024_Taentzer_HowDefineQuality}.
Moreover, these characteristics are typically defined only for \acrshortpl{IM}, as per our definitions in Section~\ref{sec:background}. In most cases, the intended use of the models to which they refer is not explicitly stated.

\subsubsection{Conceptual model evaluation frameworks}
\label{subsubsec:EvalFrame}

In addition to the specification of \glspl{QC}, the literature defines conceptual model evaluation frameworks that support the understanding and assessment of \textit{conceptual model quality}. 
This literature is closely related to the literature on \glspl{QC} reviewed in Section~\ref{subsubsec:QualCrit}. 
We classify the frameworks into three categories based on their underlying approaches.

The first category is theory-driven. 
The framework introduced by~\citet{1994_Lindland_UnderstandingQualityConceptual} aims to identify \glspl{QC} referred to as \textit{quality goals}, as well as the means of achieving them. It applies to conceptual models and the modelling process.  
\citet{2013_Krogstie_QualityConceptualData} extends this framework by incorporating principles from semiotic theory, maintaining the distinction between goals and means. It applies to the quality of models and modelling languages. 
Similarly, \citet{2012_Nelson_ConceptualModelingQuality} builds on the framework of~\citet{1994_Lindland_UnderstandingQualityConceptual} and its subsequent extensions, combining it with the ontological model of~\citet{1990_Wand_OntologicalModelInformation}. It applies to the evaluation of conceptual models and the modelling process.

The second category is practice-driven.
\citet{1994_Moody_WhatMakesGood} propose a framework for the evaluation and improvement of \acrshortpl{IM} (see definition in Section~\ref{sec:background}).  \citet{2003_Moody_ImprovingQualityData} and \citet{2003_Moody_MeasuringQualityData} later refine this framework.
Similar to the theory-driven frameworks \cite{1994_Lindland_UnderstandingQualityConceptual, 2013_Krogstie_QualityConceptualData}, they also distinguish between goals (i.e. quality factors) and means (i.e. improvement strategy). However, concepts such as \textit{stakeholders, quality review, quality issue, and improvement strategy} emphasise their practical orientation.

The third category combines both theory and practice. \citet{1997_Shanks_QualityConceptualModelling} build upon the theory-driven~\cite{1995_Krogstie_DeeperUnderstandingQuality} and the practice-driven~\cite{ 1994_Moody_WhatMakesGood} frameworks to analyse their mutual influence. 
\citet{2009_Mehmood_DataQualityConceptual} propose a meta-model for evaluating and improving the quality of conceptual models. It combines conceptual foundations with practitioner feedback. 

To summarise, all frameworks contribute to the understanding of \textit{conceptual model quality}. Most distinguish between goals and means. While theory-driven approaches focus primarily on the definition of model \glspl{QC}, the others also address model adjustments.
However, there is a lack of a framework providing a structured guiding process for the conceptual evaluation of models under development. 

\subsection{Implicit model evaluation approaches} 
\label{subsec:TestingModel}

Implicit model evaluation approaches aim to test model implementations, and as a by-product also evaluate models~\cite{2012_ETSI_MethodsTestingSpecification}. 
These approaches are applicable for models that are to be used in operation.
\citet{2012_ETSI_MethodsTestingSpecification} mentions conformance and interoperability testing as a method for implicit validation of standards relevant to \glspl{IDM1}.
Since interoperability is a critical topic for system interaction in smart grids~\cite{2018_Papaioannou_SmartGridInteroperability}, we focus in Section~\ref{subsubsec:conformanceTest} on conformance testing approaches (considered a prerequisite for interoperability testing), and in Section~\ref{subsubsec:interopTesting} on interoperability testing approaches. 
We synthesise approaches from the reviewed literature that take systems, software, and energy engineering perspectives. 

\subsubsection{Conformance testing}
\label{subsubsec:conformanceTest}
The aim of 
conformance (or \enquote{conformity}) testing~\cite{2016_NIST_ConformanceTesting} is to verify the correct implementation of a standard or specification~\cite{2018_Papaioannou_SmartGridInteroperability}.
It evaluates products or systems against the requirements specified in a standard or specification~\cite{2021_Gopstein_NISTFrameworkRoadmapa}, 
and is considered to be a prerequisite for interoperability testing~\cite{2018_Papaioannou_SmartGridInteroperability}. 
In the systems and software engineering domain, standardisation bodies propose methodological approaches for conformance testing. \citet{2021_Fernandez-Izquierdo_ConformanceTestingOntologies} summarise well-known approaches. For instance, the ISO/IEC 9646 standard specifies a general methodology for testing the conformance of products to Open Systems Interconnection specifications~\cite{_ISO/IEC_ISOIEC96461}. Based on this standard, \gls{ETSI} develops conformance test specifications at the European level~\cite{_ETSI_ConformanceTesting}. 
For application-independent test specifications, they specified TTCN-3, a testing and test control notation~\cite{2003_Grabowski_IntroductionTestingTest}.
For the energy domain, \citet{2024_Kung_EvolutionInteroperabilityStandards} and \citet{2025_Strasser_InteroperabilityTestingSmart} provide examples of conformity-related standards and frameworks for testing conformance to standards. One example is the ENTSO-E CGMES Conformity Assessment Framework~\cite{2017_entsoe_CGMESConformityAssessment}. It tests applications supporting the CIM standards 61970-600-1 and 61970-600-2 developed for \glspl{TSO} to facilitate data exchanges in system operations, network planning, and integrated electricity markets~\cite{2017_entsoe_CGMESConformityAssessment}.

Carrying out testing procedures requires adequate
infrastructure~\cite{2018_Papaioannou_SmartGridInteroperability}.
Test beds can provide this environment, including hardware, simulators, instrumentation and software tools ~\cite{2017_ISO/IEC/IEEE24765:2017E_ISOIECIEEE}. An example is the Interoperability Test Bed of the European Commission's DIGIT project, developed to support conformance testing of \gls{IT} systems~\cite{_EuropeanCommissionsDIGIT_InteroperabilityTestBed}.
Test beds should also reflect domain-specific requirements, such as those specified by~\citet{2018_Papaioannou_SmartGridInteroperability}.

To summarise, standardisation bodies and associations address conformance testing from both methodological and implementation perspectives, thereby contributing to the correct assessment of the functionality of implemented specifications.
However, because these approaches typically target models already in use, applying them during a model development process may introduce considerable complexity. This could potentially make early-stage evaluations resource-intensive and challenging to carry out.

\subsubsection{Interoperability testing}
\label{subsubsec:interopTesting}

The CEN-CENELEC-ETSI Smart Grid Coordination Group defines interoperability as \enquote{\textit{the ability of two or more networks, systems, devices, applications, or components to interwork, to exchange and use information in order to perform required functions}}~\cite{2018_Papaioannou_SmartGridInteroperability}.
Since achieving interoperable information exchange is a significant challenge, particularly in the context of a smart grid, there are various interoperability testing approaches. \citet{2025_Strasser_InteroperabilityTestingSmart} analyses existing approaches in terms of their challenges and requirements for harmonisation. 
In this section we provide an overview of some of the main approaches, focusing in particular on interoperability testing.

For instance, the \gls{JRC} defined the Smart Grid Interoperability Testing methodology~\cite{2018_Papaioannou_SmartGridInteroperability}. It aims to assess the ability of technological implementations in smart grid components to exchange and use information effectively. The focus is on the implementation of devices or protocols that are directly supported by the standard. 
The methodology consists of six steps, including use case specification, profile creation, design and analysis of the experiment, and the  testing itself.
To automate some of the activities, the \gls{JRC} has developed the Smart Grid Design of Interoperability Tests (SG-DoIT)~\citep{_JRCSmartElectricitySystems_SmartGridDesign}. Additionally, this web-based application stores the results of the interoperability testing process.

The IES approach to interoperability proposed by the Smart Grids Austria technology platform also foresees the storage of test specifications and test results~\cite{2022_Sauermann_WHITEPAPERSECTOR}. 
The approach is based on processes in the healthcare domain, and has been adapted for the energy sector. It addresses the interoperability of information exchange between \gls{ICT} systems. 
The three pillars of the approach include (1) the specification of integration profiles, (2) the organisation and implementation of interoperability peer-to-peer test events (so-called Connectathons), and (3) making publicly available the profiles and test results. The open-source software test bed Gazelle~\cite{2023_IHE-Europe_GazelleIHEEurope} supports the management of both interoperability and conformance tests, and can be used to conduct these tests.

The \gls{NIST} Framework and Roadmap for Smart Grid Interoperability~\cite{2021_Gopstein_NISTFrameworkRoadmapa} defines interoperability profiles as a subset of implementation requirements, with these derived from existing standards tailored to specific implementation. These profiles clarify which elements of a standard should be used to ensure consistent interoperability across devices and systems. They aim to reduce the complexity of implementation and testing. The framework also emphasises the importance of public accessibility and the visibility of interoperability profiles. 
Similarly, \citet{2024_Kung_EvolutionInteroperabilityStandards} identify standards for interoperability profiles within the energy domain.

Whereas further approaches (such as the ERIGrid Holistic Test Description (HTD)~\cite{2019_Heussen_ERIGridHolisticTest} or a testing methodology to check compliance with the \gls{CoC} for smart appliances) can be associated with interoperability testing, they do not address this directly.

To summarise, various approaches to interoperability testing exist, comprising methodologies, frameworks, and tools. 
Interoperability testing procedures can be synthesised into four main phases~\cite{2025_Strasser_InteroperabilityTestingSmart}: 1) identify and define test cases, 2) plan tests and set-ups, 3) execute the tests, 4) report the test results.
However, the testing approaches are not harmonised, with diverse profiling and test case specifications, as well as approaches being purpose-specific~\cite{2025_Strasser_InteroperabilityTestingSmart}.  Additionally, similarly to conformance testing procedures, interoperability testing approaches address models already in use, with execution potentially aiming for certification~\cite{2021_Gopstein_NISTFrameworkRoadmapa}. 
This indicates that applying such approaches during the development process could introduce substantial complexity, thus potentially making early-stage evaluations resource-intensive and challenging.

\subsection{Summary and research gap}
\label{subsec:SUM_GAP}

Our combined literature review provides a broad perspective on the evaluation of \glspl{IDM1} in the interdisciplinary field of energy informatics. The approaches, frameworks, and tools discussed each address specific purposes in the evaluation context.

However, the following research gaps become apparent for the evaluation of \glspl{IDM1} used in smart grid interactions.
First, although the broader literature on standards evaluation relevant for \glspl{IDM1} suggests that combining explicit and implicit methods can improve model assessment~\citep{2012_ETSI_MethodsTestingSpecification}, existing evaluation approaches focus either on implicit or explicit validation.  
Second, existing explicit evaluation approaches often remain at a high level of abstraction and typically do not provide structured, step-by-step guidance that model developers can follow. 
Third, while there is extensive literature on explicit evaluation approaches for defining and (less frequently) on evaluating the quality of conceptual models, similar work for \acrshortpl{DM} is lacking. 
Fourth, implicit validation approaches tend to focus on models that are already in use, which are often associated with formalised processes that can be excessive for evaluation of models in their development process. 
Overall, there is currently no approach that integrates both explicit and implicit validation in a structured, actionable way for use during the development process in the smart grid context, applicable for both \glspl{IDM1}.

\section{Research approach}
\label{sec:researchApproach}

Based on the gap in the research we have identified, we propose a novel evaluation method for \glspl{IDM1} used in smart grid interactions, with particular focus on the development process. We describe how we designed our method in Section~\ref{subsec:RA_evalAppr}. 
During this design process, we identified the need to consolidate sets of \glspl{QC} to support the application of our new evaluation method. We describe this consolidation process in Section~\ref{subsec:RA_QC}.

\subsection{How did we design our new evaluation method?}
\label{subsec:RA_evalAppr}
\begin{figure*}[h!]
  \centering
  \includegraphics[width=\textwidth]{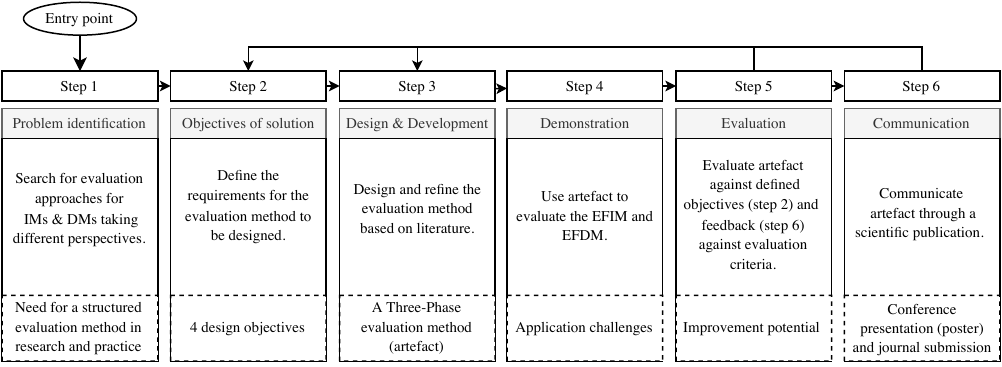}
  \caption{Adopted and applied \gls{DSRM} process based on \citet{2007_Peffers_DesignScienceResearchc}.}
  \label{fig:researchApproach}
\end{figure*}

Since we are designing a new evaluation method, we draw on the \gls{DSR} approach~\cite{2007_Peffers_DesignScienceResearchc}. Our method combines explicit model validation (to ensure \textit{conceptual model quality}) and implicit model validation (to ensure \textit{system functionality}, specifically in terms of conformance or interoperability).
The \gls{DSRM} that we follow consists of an iterative process with six steps (see Fig.~\ref{fig:researchApproach}). 

In the first step of the design process, we analyse existing evaluation approaches in order to better understand the current status of existing practice and knowledge.
To identify relevant literature, we conducted a combined literature review, as described in Section~\ref{subsec:LitReview}. 
Existing approaches often lack a combination of explicit and implicit model validation. They tend to fall short in providing a structured procedural description with actionable steps, are not tailored for \gls{DM} evaluation, and often do not account for different stages of model development. As a result, these approaches are not directly applicable and remain insufficient for evaluating \glspl{IDM1} during their development process in a smart grid interaction context. 

In the second step of the design process, insights from our literature review helped us to define the objectives for the proposed solution (i.e. our novel evaluation method).
We defined four objectives that our evaluation method should fulfil:
(1) a combination of explicit and implicit validation.
(2) suitability for evaluation practices.
(3) addresses the evaluation of both \glspl{IDM1}.
(4) adaptable to various stages of model development. 

The third step of the design process is the creation of the artefact (i.e. our proposed evaluation method). We designed our evaluation method iteratively based on the state of the art approaches to model evaluation, validation and testing (see Section \ref{sec:ModelQualEval}). 
In particular, we used the suggestions of~\citet{2012_ETSI_MethodsTestingSpecification} to combine explicit and implicit validation approaches. For the design of the explicit model evaluation phase, we used the framework proposed by~\citet{2003_Moody_ImprovingQualityData} and~\citet{2003_Moody_MeasuringQualityData} as a basis to ensure \textit{conceptual model quality} given its practical relevance (see Section~\ref{subsubsec:EvalFrame}). 
We combined it with a basic review process, as specified by~\citet{2012_ETSI_MethodsTestingSpecification}. For the design of the implicit model evaluation phase, we used the findings of the analysed literature on testing model implementations (see Section~\ref{subsec:TestingModel}) 
to ensure \textit{system functionality}, specifically in terms of conformance or interoperability. 
In order to make our evaluation approach adaptable to various stages of model development, we decided to include a modular step in the implicit model evaluation phase. This step can be replaced by sub-steps specified in selected test methods. Taking this modular approach also leads to increased compatibility of our evaluation method with various existing testing approaches and tools. It has the further advantage of maintaining the generalist nature of our evaluation method, while also ensuring its suitability for use in the smart grid domain. 

The fourth step of the design process demonstrates the use of the artefact (i.e. our proposed evaluation method) to solve the identified problem. We did this by applying it during the development of the \gls{EFIM} and \gls{EFDM}. 

The fifth step of the design process is the evaluation of the artefact (i.e. our proposed evaluation method). We conducted this evaluation based on the four objectives we defined in the second step. In addition, we gathered feedback from experts with experience in the design and use of \glspl{IDM1} across the energy informatics domain. We evaluated their feedback against three criteria defined in the Method Evaluation Model~\citep{2003_Moody_MethodEvaluationModel}: perceived ease of use; perceived usefulness; and intention to use.
Based on the results obtained, we returned to the third step and refined our evaluation approach. In total, we conducted four design iterations. 

The sixth step of the design process is communicating to relevant audiences about the artefact (i.e. our proposed evaluation method). We presented our evaluation method to international expert groups (i.e. at a symposium and a conference) and we present the final version of our evaluation method through publication in this journal.

\subsection{How did we consolidate sets of \texorpdfstring{\glspl{QC}}{QCs} for practical application?}
\label{subsec:RA_QC}

During the iterative design of our evaluation method, we identified the need to consolidate a supportive list of \glspl{QC} for the explicit \gls{IDM1} evaluation at the conceptual level.
In addition to the lack of consensus identified in the literature review (see Section~\ref{subsubsec:QualCrit}) regarding which \glspl{QC} to use, the demonstration in step 4 as part of the \gls{DSR} process revealed that existing sets of \glspl{QC} were indistinct and difficult to interpret. Moreover, not all identified defects could be mapped to the \glspl{QC} from the reviewed literature. 
To address this, we consolidated a set of practically applicable \glspl{QC} through four stages.

First, we identified 29 \glspl{QC} from the reviewed literature and consolidated them to 18 on the basis of overlapping names and concepts. 

Second, following an observation-based approach~\citep{2005_Moody_TheoreticalPracticalIssues}, we created a classification of defects based on the evaluation results of the \gls{EFIM} and \gls{EFDM}. This led to six inductively derived \glspl{QC}: \textit{semantic correctness}; \textit{usability enhancement}; 
\textit{instance uniqueness}; \textit{essentialness}; \textit{unambiguity}; and \textit{singularity}.

Third, we merged both sets, derived from the reviewed literature and the observation-based approach. This resulted in a final set of 21 \glspl{QC}, including three characteristics not previously discussed in the literature.

Fourth, we analysed this consolidated set of \glspl{QC} for \glspl{IM} to identify those that are also relevant for the evaluation of \glspl{DM}. This mapping was guided by our understanding of whether each \gls{QC} requires verification of the implementation of definitions from the \gls{IM} in the \gls{DM}, to ensure alignment between conceptual intent, and for model representation purposes. 
For example, \textit{precision}, which refers to the accuracy of attribute values required in the domain or application, is particularly relevant for the evaluation of \glspl{DM}. Verifying that the number of decimal places specified in the \gls{IM} is correctly implemented in the \gls{DM} ensures compliance with this characteristic. By contrast, \textit{naturalness}, defined as the requirement that all entities and attributes have real-world counterparts, is assumed to remain unchanged during implementation. Once this characteristic is satisfied in the \gls{IM}, it does not require re-verification in the \gls{DM}. Therefore, \textit{naturalness} is not considered applicable for \gls{DM} evaluation. 

We list our consolidated set of \glspl{QC} in Table~\ref{tab:qualityFactors} in Section~\ref{sec:evalApproach}, and we describe those we have newly proposed, in Section~\ref{sec:QC}.

\section{Description and demonstration of our artefact: A three-phase method for evaluating information and data models during their development process}
\label{sec:evalApproach}

The artefact resulting from the design process described in Section~\ref{sec:researchApproach} using \gls{DSR} is a structured evaluation method consisting of three phases.
In the subsequent description of our method, we use the design method elements outlined by~\citet{2022_Gericke_ElementsDesignMethod}.
Accordingly, we detail its core idea in Section~\ref{subsec:CoreIdea}, its intended use in Section~\ref{subsec:IntendedUse}, its representation in Section~\ref{subsec:Representation} and its procedure in Section~\ref{subsec:procedure}. The procedure description includes a demonstration of the use of our evaluation method applied to the \gls{EFIM} and \gls{EFDM}.

\subsection{Core idea of our evaluation method}
\label{subsec:CoreIdea}

The core idea of our three-phase evaluation method is to provide a structured procedure description for the evaluation of \glspl{IDM1} ensuring \textit{conceptual model quality} and \textit{system functionality}. 
We adopt an interdisciplinary approach and synthesise evaluation approaches from various disciplines of energy informatics, such as systems, software and energy engineering~\citep{2015_USLAR_EnergyInformaticsDefinition}.

Building on these different evaluation approaches, our method
integrates them into a structured description of sequential steps and decision points, enabling an iterative and practically applicable evaluation process. The formulation of the steps and decision points should be applicable across diverse use cases, ensuring applicability beyond a single model. Consequently, our method aims to support the evaluation of various \glspl{IDM1} intended to be used in smart grid interactions.

The implementation of smart grid interactions requires the design and use of both \gls{IDM1} (see Section~\ref{subsec:background_infExch}).
Therefore, our evaluation method considers the evaluation of both model types, and unlike existing approaches in the smart grid domain, also combines phases for both implicit and explicit evaluation.

Furthermore, our evaluation method is designed for use throughout the different stages of model and project development. Its modular structure allows for the adaptation of the test method so that model developers can apply our method throughout the entire model development cycle.

\subsection{Intended use of our evaluation method}
\label{subsec:IntendedUse}

Our evaluation method is intended to support model developers in systematically evaluating \glspl{IDM1} during the model development process. It targets models used in system interactions designed for automated exchange of information. In the smart grid domain, standardised models exist to describe the information to be exchanged. However, the ongoing transformation of the energy system introduces new interactions and new information to be exchanged, factors that are not always addressed by existing models.
Consequently, new models are being developed, specifically for the exchange of flexibility information. For example, the provision of \gls{DR} requires bidirectional information flows between flexibility providers and flexibility requesters, which has led to the emergence of models such as openADR and EEBus SPINE. Given the ongoing technological and regulatory transformation of smart grids, the adaptation and development of new \glspl{IDM1} are expected to remain necessary.

To ensure broad applicability across diverse smart grid use cases, we structure our method at a high level. However, for the execution of evaluations, model developers are encouraged to complement our method with domain-specific methods and tools, such as described in Section~\ref{sec:ModelQualEval}. This allows for more specific evaluation procedures. 

Our method can be applied at various stages of model development. The procedure description distinguishes between formal and informal test types. Informal test types should allow evaluations to be made in early model development stages. The earlier defects are identified, the lower are the correction costs in comparison to corrections at a later development stage~\citep{2003_Moody_ImprovingQualityData}. The application of our method therefore has the benefit of increasing cost-effective model development, and also increases the models' reliability when deployed. In addition, it enables the comparability of models through a common evaluation approach and the use of defined \glspl{QC} that can support certification efforts.

Overall, we do not assume any specific knowledge for the application of our evaluation method. However, it is assumed that all stakeholders involved in the evaluation have the same understanding of the \glspl{QC} used.

\subsection{Representation of our evaluation method}
\label{subsec:Representation}
\begin{figure*}[ht!]
  \centering
  \includegraphics[width=0.85\textwidth]
  {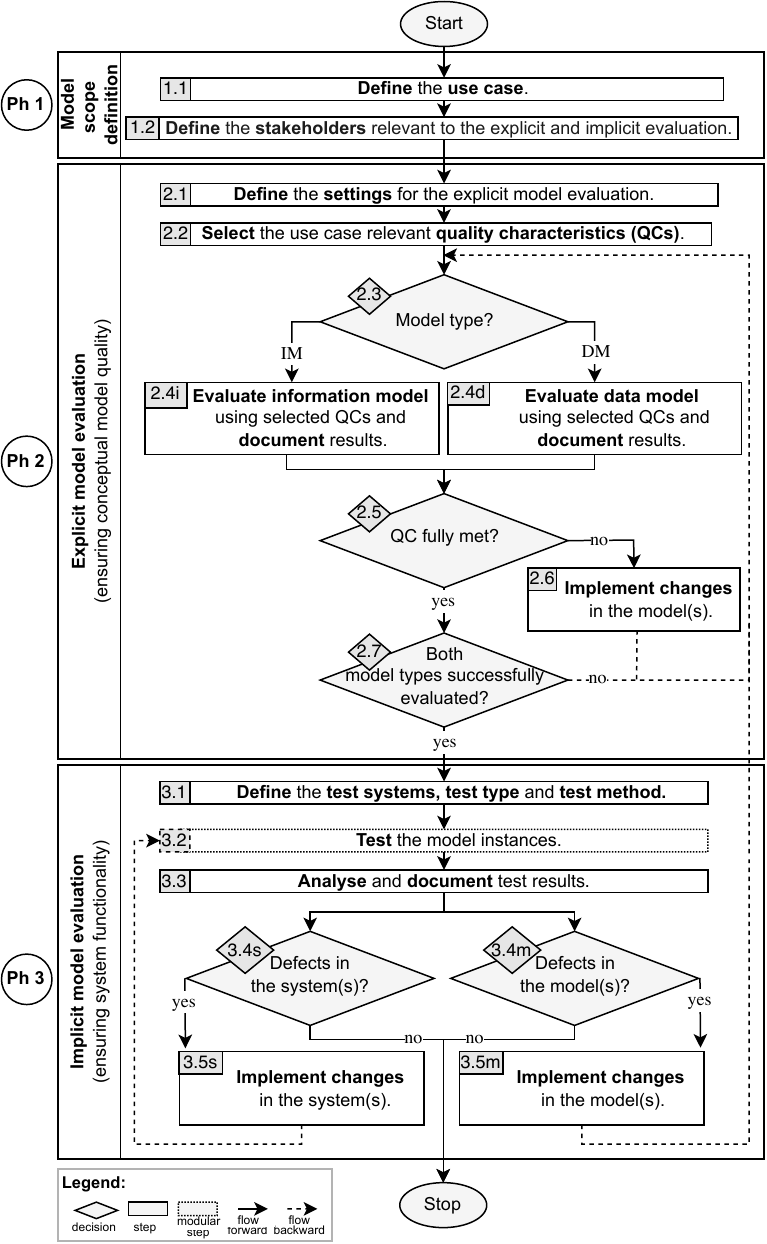}
  \caption{Three-Phase Evaluation Method for \glspl{IDM1} in their development process.}
  \label{fig:evalApproach}
\end{figure*}

To support comprehension and reproducibility, we illustrate our evaluation method as a structured flow chart in Fig.~\ref{fig:evalApproach}. This visual representation highlights the systematic nature of our evaluation method, including sequential steps to be followed.
The steps are grouped into three phases: 
Phase 1 - model scope definition; Phase 2 - explicit model evaluation; and Phase 3 - implicit model evaluation. 
Since our method describes an iterative evaluation process, phases might be entered by different steps.
We use the following visual conventions:
\begin{itemize}
    \item rectangles represent the core steps of the evaluation procedure;
    \item dotted rectangles indicate the modular character of a step, allowing for the integration of domain-specific methods;
    \item Trapezoids represent decision points that guide the flow based on outcomes of a previous step;
    \item continuous arrows indicate the logical progression from one step to the next;
    \item dashed arrows represent iterative loops, where steps may be revisited based on a previous decision.
\end{itemize}

\subsection{Procedure description and example demonstration of our evaluation method}
\label{subsec:procedure}

We introduce the models – which serve as the basis for demonstrating our evaluation method – in Section~\ref{subsubsec:EFIM_EFDM}. Each subsequent section describes one phase of our method, following a consistent structure: a procedure description outlining the phase’s purpose and steps, followed by a demonstration applying our method to the models. The procedure descriptions also provide guidance for applying our method in smart grid-specific contexts.

\subsubsection{Information and data models used for example demonstration}
\label{subsubsec:EFIM_EFDM}

The \gls{EFIM} and \gls{EFDM} have been developed as part of the German research project SynErgie that addresses the flexible alignment of industrial processes to fluctuating power supply~\citep{SynErgie}. The models define the descriptions of industrial energy flexibility potential and related load measures~\cite{2023_Lindner_EnergyFlexibilityDataa}. One key feature is their applicability across various industrial sectors, that reduces interoperability challenges associated with information exchange~\cite{2025_vanStiphoudt_EnergySynchronizationPlatform}. The research and industry experts involved in the model development process are specialised in the energy sector, manufacturing, and \gls{IT}. 

The \gls{EFIM} and \gls{EFDM} consist of two main classes. The \texttt{flexibilitySpace} class describes the potential (i.e. possible options) of an industrial company to adjust its planned power consumption. The \texttt{flexibleLoadMeasuresPackage} class describes the schedule (i.e. one specific option) on which an industrial company adjusts its planned power consumption. The \gls{EFIM} is specified in German and English, and the \gls{EFDM} in \gls{JSON} schemas (i.e. one schema for each class). A complete model specification can be found in a Git repository~\cite{2023_Lindner_EnergyFlexibilityDataa}. Overall, version 1.0 of these models comprises 32 main elements.

Using these models in information exchanges (for example, to enable the provision of industrial \gls{DR}), leads to \gls{EFDM} instances being exchanged between an industrial company and an energy service provider. Listing~\ref{lst:EFDM} illustrates a JSON snippet of such an example of an \gls{EFDM} instance being exchanged, in which potential power states for a flexible load are specified, for example, as is the case with industrial machinery.

\begin{minipage}{\textwidth}%

\lstset{
    frame=single,
    framesep=5pt,
    xleftmargin=0pt,
    xrightmargin=0pt,
    framexleftmargin=10pt,
    framexrightmargin=-31pt,
    linewidth=\linewidth,
    language=json,
    breaklines=true,
    basicstyle=\ttfamily\footnotesize,
}
\begin{lstlisting}[caption={Snippet of a \texttt{flexibilitySpace} instance (adapted from~\citep{2023_Lindner_EnergyFlexibilityDataa}).}, label={lst:EFDM}]
{   ...

    "origin": {
          "originId": "6e8bc430-9c3a-11d9-9669-0800200c9a67",
          "timestamp": "2023-10-26T14:30:15Z"
        },
    ...
    
     "flexibleLoads": [
      {
        "flexibleLoadId": {
          "uuid": "6e8bc430-9c3a-11d9-9669-0800200c9a69",
          "comment": "Flexible load UUID"
        },
    ...
    
        "powerStates": {
              "order": "chronological",
              "durationType": "deliveryDuration",
              "values": [
                {
                  "power": {
                    "unit": "kW",
                    "minValue": 1.000,
                    "maxValue": 5.000
                  },
                  "duration": {
                    "unit": "s",
                    "minValue": 1.001,
                    "maxValue": 10.000
                  }
                }
              ]
            },
    ...
\end{lstlisting}
\end{minipage}

\subsubsection{Phase 1 - Model scope definition}
\label{subsubsec:Ph1}
\paragraph{Procedure description}
As illustrated in Fig.~\ref{fig:evalApproach}, this phase consists of two steps. In step 1.1, the model developer(s) define the intended use case for the model being evaluated. Use cases establish a shared understanding among stakeholders, which is crucial in complex systems such as smart grids~\cite{2018_Papaioannou_SmartGridInteroperability, 2023_PotencianoMenci_SYSTEMORGANIZATIONOPERATION}.
They provide the basis for selecting suitable \glspl{QC} in Phase 2 and defining the test cases
~\cite{2018_Papaioannou_SmartGridInteroperability} in Phase 3.
Use case templates, such as the one defined in IEC 62559-2, provide a uniform specification~\cite{2024_IEC_IEC62559Use}. Their use reduces effort and enables comparability between descriptions.
In the context of smart grids, the standardisation bodies CEN, CENELEC and ETSI provide a checklist aligned with IEC 62559-2. This checklist (which consists mainly of binary questions) is intended to facilitate the description of use cases that align with the \gls{SGAM}~\cite{2014_CEN_SGAMUserManual}.
Since our evaluation method is intended for the evaluation of \glspl{IDM1}, (which are to be used specifically in system interactions), certain fields in this checklist are identical for all use cases.
In particular, the \textit{nature of use cases} is always technical and focuses on the specification of information to be exchanged. The \textit{viewpoint} is aligned with the \gls{SGAM} information layer, and the \textit{scope} is limited to the information exchange specified according to the \gls{IDM1} under evaluation. Potential hierarchical links to broader use cases can be indicated within the field 
\textit{relation to other use cases}. All other fields are use case-specific.  

In step 1.2, after the use case specification, the model developer(s) identify participants for the evaluation process. They select individuals from the stakeholder groups identified in the previous step. The involvement of stakeholders ensures that the process takes account of various perspectives and interests during the evaluation process~\cite{2003_Moody_ImprovingQualityData, 2003_Moody_MeasuringQualityData}. 
Afterwards, the model developer(s) and stakeholder(s) proceed with the explicit model evaluation in Phase 2.

\paragraph{Example demonstration}

We provide two use case descriptions in Table~\ref{tab:useCases}, based on the template specification by~\citet{2014_CEN_SGAMUserManual}.
These use cases are derived from system interactions in the \gls{ESP}, a digital platform concept designed to streamline automated industrial ~\gls{DR}~\citep{2025_vanStiphoudt_EnergySynchronizationPlatform}.
The flexibility information exchanged in the \gls{ESP} is described using the \gls{EFIM} and \gls{EFDM}, which serve as models to be evaluated in our example demonstrations. 

The identical fields of the filled template in Table~\ref{tab:useCases} for both use cases are their technical nature, the \gls{SGAM} information layer as a viewpoint, and the scope of information exchange described by the \gls{EFIM} and \gls{EFDM}.  
Additionally, both use cases relate to the \gls{SGAM} domains of customer premises and span across the \gls{SGAM} zones market, enterprise and operation.

\begin{figure}[!ht]
  \centering
    \includegraphics[width=0.8\textwidth]{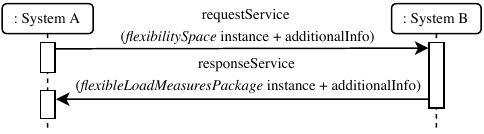}
    \caption{Sequence diagram of the information exchange of the first use case \textit{assess flexibility}.}
    \label{fig:sequence1}
\end{figure}

The first use case, \textit{assess flexibility}, involves two systems: the industrial company's information system (System A) and the flexibility assessment service provider's \gls{API} (System B). Fig.~\ref{fig:sequence1} illustrates the sequence diagram for this interaction. 

The objective is to enable successful information exchange that supports the company’s decision-making on when, where, and how to activate industrial flexibility in an economically efficient manner. 
To achieve this, the industrial company describes its flexibility potential using the \texttt{flexibilitySpace} class of the \gls{EFDM}. This class specifies entities and attributes such as power states, holding duration, activation and regeneration times, activation/deactivation gradients.
Additional information, such as the period under consideration and the relevant electricity markets, are sent together with the \texttt{flexibilitySpace} instance to the flexibility assessment service provider's \gls{API}. Based on this information, the service provider then computes an optimal flexibility activation schedule and returns results, including
expected revenue, associated costs, energy consumption, and the applicable electricity markets. The resulting flexibility schedule is described as load profiles using the \texttt{fexibleLaodMeasuresPackage} class of the \gls{EFDM}.

\begin{table}[!ht]
\centering
\caption{System use case description of example applications based on use case fields from~\citet{2014_CEN_SGAMUserManual}.}
\label{tab:useCases}
\small
\begin{tabular}{p{0.22\textwidth}p{0.33\textwidth}p{0.33\textwidth}}
\toprule
\textbf{Use case} & 
    \textbf{Use case 1} & 
    \textbf{Use case 2} \\ 
\midrule

    Name & 
        Assess Flexibility & 
        Provide Flexibility \\

\midrule
    \gls{SGAM} Domains/Zones &
        \multicolumn{2}{c}{Customer Premises/ Market, Enterprise, Operation}
        \\
\midrule
    Narrative &
        The industrial company sends its flexibility potential to an external service provider to assess the economic value and obtain the best schedule. &
        The industrial company sends its flexibility potential to a local flexibility market service provider to provide \gls{DR}.\\
\midrule
    Actors (systems) & 
        Information systems of an industrial company and a flexibility assessment service provider &
        Information systems of an industrial company and a local flexibility market service provider \\

\midrule
    Nature of use case &
        \multicolumn{2}{c}{Technical (information)}
       \\
\midrule
    Viewpoint &
        \multicolumn{2}{c}{\gls{SGAM} Information layer}
       \\
\midrule
    Objective & 
        Successful interaction between the two actors (systems) to support decision-making on when, where, and how to activate industrial flexibility in an economically efficient manner. &
        Successful interaction between the two actors (systems) to support the matching of actors and their flexibility potential and needs. 
        
    \\
\midrule
    Scope &
    \multicolumn{2}{c}{Information exchange using the \gls{EFIM} and \gls{EFDM}.} 
    \\
\midrule

    Information exchanged & 
        flexibility potential (power states, holding duration, usage number, activation and deactivation gradients, costs per activation, prices, regeneration time), electricity markets, dependencies (target and triggering flexibility, timely reference points, dependency types, dependency periods); load profile, costs, savings, energy consumed, electricity markets & 
        flexibility potential (power states, holding duration, usage number, activation and deactivation gradients, costs per activation, prices, regeneration time), dependencies (target and triggering flexibility, timely reference points, dependency types, dependency periods); load profile, revenue\\

\bottomrule
\end{tabular}%
\end{table}

\begin{figure}[!ht]
  \centering
    \includegraphics[width=1\textwidth]{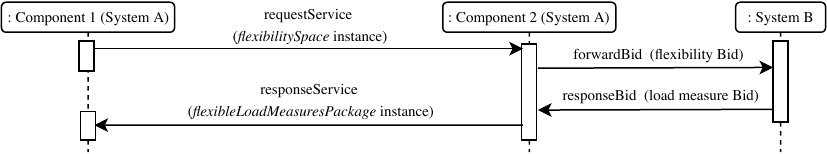}
    \caption{Sequence diagram of the information exchange of the second use case \textit{provide flexibility}.}
    \label{fig:sequence2}
\end{figure}

The second use case, \textit{provide flexibility}, involves two systems: the industrial company’s digital company-side platform (System A), comprising two internal components, and a local flexibility market service provider's \gls{API} (System B). Fig. ~\ref{fig:sequence2} illustrates the sequence diagram for this interaction. 

The objective is to enable information exchange for the provision of \gls{DR}. To achieve this, the industrial company first describes its flexibility potential using the \texttt{flexibilitySpace} class of the \gls{EFDM}, similar to the first use case. Afterwards, component 1 sends this instance to component 2 of the digital company-side platform. Component 2 then transforms this information into the local flexibility market's data format and sends it as a flexibility bid to the local flexibility market. The service provider matches the flexibility bid with offers from \glspl{DSO} that act as flexibility requesters and sends back a reservation or activation message to component 2. This message is transformed into an \gls{EFDM} class \texttt{fexibleLaodMeasuresPackage} instance, which includes details such as load profiles and expected revenue. Finally, component 2 forwards this information to component 1, enabling activation signals to be sent to machines according to the agreed load profiles. 

In addition to the use case description, Phase 1 of our evaluation method includes a step for the specification of the involved stakeholders, relevant for the two subsequent phases.
For the evaluation of the \gls{EFIM} and \gls{EFDM} in the two use cases, three additional stakeholders (i.e. developers of the energy service and experts in manufacturing and \gls{IT}), are included alongside the model developers.

\subsubsection{Phase 2 - Explicit model evaluation}
\label{subsubsec:Ph2}
\paragraph{Procedure description}
As illustrated in Fig.~\ref{fig:evalApproach}, this phase consists of five steps that define a review process for evaluating the \textit{conceptual model quality} of the \glspl{IDM1} under evaluation. 
Different types of reviews are applicable in this context, notably inspection~\cite{2003_Moody_ImprovingQualityData, 2012_ETSI_MethodsTestingSpecification, 2009_Mohagheghi_DefinitionsApproachesModel} and walk-throughs~\cite{1999_ETSI_MethodsTestingSpecification, 2012_ETSI_MethodsTestingSpecification} are mentioned in the context of model/standard evaluation.

In step 2.1 of our evaluation method, model developer(s) define the settings for the explicit model evaluation, including planning and preparation of the review process.
This may involve appointing a chair that leads the review meetings, inviting reviewers, scheduling review dates, and providing the models or specifications to reviewers prior to the review meetings~\citep{1999_ETSI_MethodsTestingSpecification, 2012_ETSI_MethodsTestingSpecification}. Reviewers may also be asked to review the models or specifications in advance~\citep{2012_ETSI_MethodsTestingSpecification}. In line with this,
~\citet{2003_Moody_ImprovingQualityData} recommends rating the models for each \gls{QC} on a five-point scale prior to the review meetings. 

In addition to the organisational planning and preparation, the model developer(s) select the \glspl{QC} in step 2.2 relevant to the use case defined in phase 1.
We provide an overview of a consolidated set of \glspl{QC} in Table~\ref{tab:qualityFactors}. While all \glspl{QC} are relevant for the evaluation of \glspl{IM}, only 12 are relevant for the evaluation of \glspl{DM}.

\begin{ThreePartTable}
\begin{small}

\begin{TableNotes}
\footnotesize
  \item[1] Defect observed.
  \item[2] Analytical explanation.
  \item[3] OBA being observation-based approach.
\end{TableNotes}

\begin{longtable}{p{0.16\textwidth}p{0.055\textwidth}p{0.28\textwidth}p{0.15\textwidth}p{0.2\textwidth}}
\caption{Supportive list of \gls{QC} for the evaluation of \glspl{IM} and \glspl{DM}.}
\label{tab:qualityFactors}\\

\toprule
\gls{QC} & Model type & Evaluation question & Observed defect example & Based on referenced \gls{QC} \\
\midrule
\endfirsthead

\multicolumn{5}{l}{\textit{Table \thetable\ (continued)}}\\
\toprule
\gls{QC} & Model type & Evaluation question & Observed defect example & Based on referenced \gls{QC} \\
\midrule
\endhead

\bottomrule
\insertTableNotes
\endfoot

Semantic correctness
  & \gls{IM}
  & Does the model correctly specify entities/ attributes for the intended purpose?
  & Attributes with incorrect unit.
  & relevance~\cite{1995_Levitin_QualityDimensionsConceptual};
    \newline semantic correctness [OBA]\tnote{3} \\

Completeness
  & \gls{IM}/ \gls{DM}\tnote{2}
  & Does the model consider all entities/attributes required for the intended application?
  & Entity missing.
  & completeness~\cite{2003_Moody_ImprovingQualityData}; 
    \newline complete~\cite{1999_Lee_InformationModelingDesign}
    \newline comprehensive\-ness~\cite{1995_Levitin_QualityDimensionsConceptual}\\

Unambiguous\-ness
  & \gls{IM} 
  & Are all entities/ attributes clear and without the possibility of misinterpretation?
  & Definitions of entities unclear.
  & unambiguous definitions~\cite{1995_Levitin_QualityDimensionsConceptual};
    \newline unambiguous~\cite{1999_Lee_InformationModelingDesign};
    \newline unambiguity [OBA]\tnote{3}\\
     
Understand\-ability
  & \gls{IM}/ \gls{DM}\tnote{1}
  & Are all entities/attributes easily understandable?
  & Data type not human-readable.
  & understand\-ability~\cite{2003_Moody_ImprovingQualityData};
    \newline increase usability [OBA]\tnote{3}\\

Value obtainability
  & \gls{IM}
  & Are values for all entities/attributes sufficiently accessible?
  & n/a
  & obtainability of values~\cite{1995_Levitin_QualityDimensionsConceptual}\\

Simplicity
  & \gls{IM}
  & Are only entities/attributes specified that are necessary for the intended use case?
  & n/a
  & essentialness~\cite{1995_Levitin_QualityDimensionsConceptual};
    \newline simplicity~\cite{2003_Moody_ImprovingQualityData} \\

Attribute granularity
  & \gls{IM} 
  & Is the granularity of entity/ attribute specifications achievable and justifiable?
  & Entity requires min and max attributes.
  & attribute granularity~\cite{1995_Levitin_QualityDimensionsConceptual}\\

Precision
  & \gls{IM}/ \gls{DM}\tnote{2}
  & Does the specified accuracy of the attribute values fulfil the requirements of the domain/application?
  & Incorrect number of decimals for attributes.
  & domain precision~\cite{1995_Levitin_QualityDimensionsConceptual};
    \newline precise~\cite{1999_Lee_InformationModelingDesign}\\

Naturalness
  & \gls{IM}
  & Do all entities/attributes have a real counterpart, and do attributes represent a single fact?
  & n/a
  & naturalness~\cite{1995_Levitin_QualityDimensionsConceptual}\\

Identifiability
  & \gls{IM}/ \gls{DM}\tnote{2}
  & Are individual entities/attributes identifiable and distinct from one another?
  & Identical attribute names and enumerations. 
  & occurrence identifiability~\cite{1995_Levitin_QualityDimensionsConceptual};
    \newline sharable~\cite{1999_Lee_InformationModelingDesign} \\
  
Homogeneity 
  & \gls{IM}/ \gls{DM}\tnote{2}
  & Do all attributes of a given type apply to all entities of that type?
  & Two attributes define variants of an attribute. 
  & homogeneity~\cite{1995_Levitin_QualityDimensionsConceptual}\\

Semantic consistency
  & \gls{IM}/ \gls{DM}\tnote{2}
  & Are all entities/attributes consistently defined between related components and across related models?
  & n/a
  & semantic consistency~\cite{1995_Levitin_QualityDimensionsConceptual};
    \newline integration~\cite{2003_Moody_ImprovingQualityData}\\

Structural consistency
  & \gls{IM}/ \gls{DM}\tnote{2}
  & Are entities/attributes consistently structured?
  & Inconsistent hierarchical structure of attributes.
  & structural consistency~\cite{1995_Levitin_QualityDimensionsConceptual}; 
    \newline well-structured~\cite{1999_Lee_InformationModelingDesign}\\

Robustness
  & \gls{IM}/ \gls{DM}\tnote{2}
  & Are all entities/ attributes stable against changing requirements?
  & n/a
  & robustness~\cite{1995_Levitin_QualityDimensionsConceptual};
    \newline stable~\cite{1999_Lee_InformationModelingDesign}\\

Extensibility
  & \gls{IM}/ \gls{DM}\tnote{2}
  & Is the model easily adaptable to incorporate changes?
  & n/a
  & flexibility~\cite{1995_Levitin_QualityDimensionsConceptual, 2003_Moody_ImprovingQualityData};
    \newline extensible~\cite{1999_Lee_InformationModelingDesign}\\

Implement\-ability
  & \gls{IM} 
  & Can the model be implemented within the project requirements?
  & not considered
  & implement\-atbility~\cite{2003_Moody_ImprovingQualityData}\\
  
Integrity
  & \gls{IM}
  & Are all business rules that apply to the data incorporated?
  & not considered
  & integrity~\cite{2003_Moody_ImprovingQualityData}\\

Syntactic correctness
  & \gls{IM}/ \gls{DM}\tnote{2}
  & Are modelling conventions met?
  & Undesirable behaviour of numeric types. 
  & correctness~\cite{2003_Moody_ImprovingQualityData}\\

\textbf{Singularity}
  & \gls{IM}/ \gls{DM}\tnote{2}
  & Does every entity/attribute describe distinct information? 
  & Redundant attributes for model version.
  & \textbf{proposed in this manuscript}\\

\textbf{Instance uniqueness}
  & \gls{IM}
  & Does the model specify an attribute to distinguish model instances?    
  & No ID attribute for model schema.
  & \textbf{proposed in this manuscript}\\

\textbf{Essentialness}
  & \gls{IM}/ \gls{DM}\tnote{2}
  & Are only entities/attributes specified as mandatory that are necessary for all intended use cases?
  & All attributes are mandatory by default.
  & \textbf{proposed in this manuscript}\\

\end{longtable}
\end{small}
\end{ThreePartTable}

Next, the model developer(s) and stakeholder(s) decide which model type to evaluate (see decision 2.3). 
Our evaluation method includes a separate assessment of \glspl{IM} and their related \glspl{DM} to ensure that each model type is adequately addressed. For \glspl{IM}, the modelling context is particularly important; for example, \textit{completeness} is evaluated based on whether the information to be exchanged is adequately represented for the intended use case. In contrast, the evaluation of \glspl{DM} focuses on  verifying that all entities and attributes defined in the \gls{IM} are properly implemented and implementation-specific details, such as data types are included.

The explicit evaluation process always begins with the evaluation of the \gls{IM}.
In step 2.4i, the model developer(s) and stakeholder(s) review the \acrshort{IM}, evaluate it against the chosen \glspl{QC} and document the results.  
The specific procedure depends on the settings defined in step 2.1.  In addition to human inspection, model developer(s) and stakeholder(s) may use tools to support defect detection~\cite{2009_Mohagheghi_DefinitionsApproachesModel}.
For the documentation of the review results, \citet{2003_Moody_ImprovingQualityData} propose a quality issues matrix. That matrix includes fields for the issue description, related \glspl{QC}, possible resolutions (i.e. changes to address the issues), priority in terms of their correction and their status (i.e. resolved, open, etc.).

Based on the evaluation and documentation, the stakeholders decide in decision 2.5 whether all \glspl{QC} are met.
If defects are identified, the model is revised in step 2.6 and changes are implemented. The evaluation starts again from decision 2.3.
This iterative process continues until all \glspl{QC} are satisfied. Once the \gls{IM} is successfully evaluated, the same process is applied to the \gls{DM} in step 2.4d.
Once both model types are successfully evaluated and all \glspl{QC} are met (see decision 2.7), the model fulfils the requirements for \textit{conceptual model quality} and the model developer(s) and stakeholder(s) will proceed with the implicit model evaluation in Phase 3.

\paragraph{Example demonstration}
For the evaluation of the \gls{EFIM} and \gls{EFDM}, the model developers and stakeholders plan and prepare the review session.
A chair is appointed to lead the session, this may be a model developer or stakeholder. While \gls{ETSI} recommends appointing a neutral chair not affiliated with model development~\citep{2012_ETSI_MethodsTestingSpecification}, resource constraints in research or smaller projects may necessitates deviations. In this demonstration, the chair is one of the model developers.

Following step 2.2, the model developers and stakeholders select 19 out of the 21 \glspl{QC} listed in Table~\ref{tab:qualityFactors} based on the use case specified in step 1.1. Since the \gls{EFIM} and \gls{EFDM} are not tailored to specific organisational applications, the \glspl{QC} \textit{integrity} and \textit{implementability}, which relate to compliance with business rules and project requirements, are excluded from the evaluation.

The initial explicit model evaluation begins with the \gls{EFIM} (see decision 5). The model is evaluated against the selected 19 \glspl{QC} through human inspection according to step 2.4i and any identified defect is documented. As the models are hosted in GitLab, issues are logged directly within the platform~\citep{GitLab_EFDM}.
If one or more \glspl{QC} are not met (see decision 2.5), the \gls{EFIM} is revised to address the documented defects and proposed changes (step 2.6).
This iterative process continues until the \gls{EFIM} satisfies all selected \glspl{QC}. The evaluation then proceeds with the \gls{EFDM}, following steps 2.4d to 2.6. 

Examples of identified defects in steps 2.4d/2.4i include missing price elements (\textit{\gls{QC} completeness}), incorrect units (\textit{\gls{QC} semantic correctness}), redundant information such as two elements referring to the model version (\textit{\gls{QC} singularity}) and absence of IDs (\textit{\gls{QC} instance uniqueness}). Once both \gls{EFIM} and \gls{EFDM} are successfully evaluated and meet all \glspl{QC} (see decision 2.7), the model fulfils the requirements for \textit{conceptual model quality} and the evaluation proceeds with Phase 3.

\subsubsection{Phase 3 - Implicit model evaluation}
\label{subsubsec:Ph3}
\paragraph{Procedure description}
As illustrated in Fig.~\ref{fig:evalApproach}, this phase consists of four steps. These define a testing procedure for model implementations aimed at validating \textit{system functionality}, specifically in terms of conformance or interoperability. This phase involves validating model implementations through testing, analysing defects and implementing defect corrections.
In step 3.1, model developer(s) and stakeholder(s) select the test system(s), test type and test method.
The test system(s) are chosen based on the models' scope defined in Phase 1. If no system has yet been implemented, model developer(s) and stakeholder(s) can either develop a test system or define a conceptual test system based on the use case description in step 1.1.
Our evaluation method differentiates between two types of test: formal and informal. 
Formal tests follow established test procedures and offer the advantage of applicability beyond the development process. For instance, the Interoperability Test Bed~\cite{_EuropeanCommissionsDIGIT_InteroperabilityTestBed}, enables the specification and provision of reusable test cases, which can later be used by external stakeholders to validate their implementations of the deployed \glspl{IDM1}.
However, the suitability of formal tests in model development projects must be assessed in terms of the model's and project's development stage, considering the
required effort and time~\cite{2012_ETSI_MethodsTestingSpecification}.  
Informal tests include simplified versions of formal test procedures or conceptual testing. Simplified tests are appropriate when test systems are not yet sufficiently mature to support automated testing. Conceptual tests are particularly suitable during early development stages.

Once the test type is selected, the model developer(s) and the stakeholder(s) define a suitable test method aligned with the previously chosen test type.
We provide an overview of test methods and test tools for conformance and interoperability testing in Section~\ref{subsec:TestingModel}. For informal tests (i.e. simplified or conceptual), the model developer(s) and stakeholder(s) have to define their own method. In the case of a conceptual test, data is mapped to the \gls{IDM1} under evaluation according to the defined conceptual test system.

In step 3.2, the model developer(s) and stakeholder(s) conduct the test according to the previously chosen test method. This step is modular and can be replaced or extended by specific sub-steps, depending on the previously selected or defined test method. This modularity is indicated in Fig.~\ref{fig:evalApproach} by a dotted line, highlighting the possibility to tailor the validation procedure to the development status of the model or project.

As the test outcome is binary, either pass or fail, and the goal of the evaluation is to improve the developed models, the model developer(s) and stakeholder(s) have to analyse and document the test results in step 3.3. In the event of a failure, this documentation serves to identify the underlying cause.

If the model developer(s) and stakeholder(s) identify defects in the system(s) (see decision 3.4s), the model developer(s) implement changes to fix the defects in step 3.5s and re-execute the test from step 3.2
If the model developer(s) and stakeholder(s) identify defects in the model(s) (see decision 3.4m), the model developer(s) revise and implement changes to the model(s) in step 3.5m and start re-evaluating the model(s) with decision 2.3 in Phase 2. This iterative process continues until no defects are identified in either the system (see step 3.4s) or the model (see step 3.4m). 
Once the model developer(s) and stakeholder(s) have successfully conducted all relevant tests, the evaluation of both \acrshort{IM} and \acrshort{DM} is deemed successful, and the model evaluation process is considered complete and concludes.

\paragraph{Example demonstration}

For the evaluation of the \gls{EFIM} and \gls{EFDM}, model developers and stakeholders select a test system in step 3.1 for each use case. For the first use case, they choose the Flex-Tool, and for the second use case, the Local flexibility market (i.e. System B in Fig.~\ref{fig:sequence1} and~\ref{fig:sequence2}).
Given the status of the research project at the time of the demonstration, the model developers and stakeholders choose the informal test type and conduct a simplified form of a conformance test.
Thus, model developers and stakeholders define a customised test method based on the formal conformity testing approach outlined by~\citet{_ETSI_ConformanceTesting}. The resulting five sub-steps replace the modular step 3.2. In the following, we detail these steps only for the first use case, as the procedure is similar for both use cases.

In step 3.2.1, the model developers and stakeholders specify the test case. Following our use case, an industrial company sends \texttt{flexibilitySpace} instances from its digital company platform (System A) to the Flex-Tool (System B) (see Fig.~\ref{fig:sequence1}). 
The expected responses consists of \texttt{flexibleLoadMeasuresPackage} instances that the Flex-Tool sends to the digital company platform.

In step 3.2.2, the model developers and stakeholders select the syntax and semantics test criteria. The syntax test criteria ensure that serialised class instances conform to the corresponding class schema specifications. The semantics test criteria ensure that the values specified in the response (i.e. \texttt{flexibleLoadMeasuresPackage} instances) sent from the Flex-Tool to the digital company platform fall with the valid value ranges (e.g. minimum and maximum values) defined in the request (i.e. \texttt{flexibilitySpace} instances).

In step 3.2.3, the model developers and stakeholders select the test environment. For the syntax test, they decide to write test scripts in Python. For the semantics test, they decide to perform a manual test execution (i.e. manually comparing the values in the response with the semantics test criteria) due to time constraints.

In step 3.2.4, the industrial company defines the \texttt{flexibilitySpace} instances and implements the information exchange in collaboration with the model developers using an \gls{API} client.

In step 3.2.5, the model developers execute the specified tests.

Defects identified after failed syntax tests are ascribed to incorrect \gls{EFDM} implementation in the test system. Defects identified after failed semantic tests resulted, for example, from specification issues, such as missing unit definitions in the \gls{EFDM}, leading to inconsistent value magnitudes.
Depending on the nature of the defect, i.e. whether system-related (decision 3.4s) or model-related (decision 3.4m), the tests are re-executed in step 3.2 or re-evaluated from decision 2.3 in Phase 2 onwards.

Once both syntax and semantics tests are successfully passed, the model evaluation process for the considered model development stage is considered complete even though the development of the models continues. The model versions before and after these evaluations are available in a public Git repository~\cite{2023_Lindner_EnergyFlexibilityDataa}.

\section{Our proposed set of consolidated model quality characteristics}
\label{sec:QC}
In Table~\ref{tab:qualityFactors}, we list the consolidated set of 21 \glspl{QC} resulting from the process described in Section~\ref{subsec:RA_QC}. This set complements the explicit model evaluation in Phase 2 of our proposed evaluation method and defines \textit{conceptual model quality}. 

We specify each \gls{QC} with an evaluation question based on the definitions and descriptions in the reviewed literature. Where defects were observed during the application of our evaluation method to the \gls{EFIM} and \gls{EFDM}, we mentioned an illustrative example to enhance practical applicability and respond to expert feedback regarding the \textit{Perceived ease of use} (see Section~\ref{sec:eval}).

Furthermore, we map the \glspl{QC} to the two model types as described in Section~\ref{subsec:RA_QC}. Whereas all 21 \glspl{QC} apply to \glspl{IM}, 12 are relevant for \glspl{DM}. The consolidated set also introduces three new characteristics: \textit{singularity}, \textit{instance uniqueness}, and \textit{essentialness}. We derived these \glspl{QC} from an observation-based approach, enabling us to address defects not adequately captured by existing \glspl{QC}. We thereby strengthen the practical relevance of our evaluation method.
As the definitions of the consolidated \glspl{QC} are provided in the referenced literature, we will focus on these three new ones in the following:

\textit{Singularity} refers to the absence of redundant elements, which helps to maintain a single point of truth and is critical to prevent the processing of contradictory information.
While~\citet{2003_Moody_ImprovingQualityData} subsumed redundancy under the \gls{QC} \textit{correctness}, this association may not be consistently recognised by non-technical stakeholders involved in the development of domain models. To mitigate the risk that this characteristic is being overlooked, we include \textit{singularity} as a distinct \gls{QC} in our proposed list.

\textit{Uniqueness} refers to the requirement that model instances have unique identifiers, ensuring that they remain distinguishable during automated exchanges, storage, and processing. By specifying identifiers directly in \acrshortpl{IM}, the need for \gls{IT} systems to generate them independently is eliminated, thereby reducing the likelihood of misidentification errors.

\textit{Essentialness} refers to the minimum specification of mandatory entities and attributes, helping to prevent the inclusion of intentionally false information in \acrshort{DM} instances. Once incorrect data are exchanged between information systems, it is often difficult to detect and correct them. To support model expandability and ensure backward compatibility with older model versions, default values may be defined for non-mandatory elements.

\section{Evaluation of our designed artefact} 
\label{sec:eval}

As stated by~\citet{2022_Gericke_ElementsDesignMethod}, validation of a method is conceptually and practically difficult. In particular, we cannot capture the success of our evaluation method by a binary or threshold value; a limitation also acknowledged by~\citet{2022_Gericke_ElementsDesignMethod} for many design methods.
To address this, we carry out a qualitative evaluation, assessing our artefact against the objectives defined as part of the applied \gls{DSRM} process model (see Section~\ref{sec:researchApproach}). 
Further, we collect expert feedback through different means and evaluate it against three criteria for the evaluation of design methods.
We take these criteria from the Method Evaluation Model~\citep{2003_Moody_MethodEvaluationModel}:
The first criterion, \textit{Perceived ease of use}, assesses \enquote{the degree to which a person believes that using a particular method would be free of effort}.
The second criterion, \textit{Perceived usefulness}, assesses \enquote{the degree to which a person believes that a particular method will be effective in achieving its intended objectives}.
The third criterion, \textit{Intention to use}, assesses \enquote{the extent to which a person intends to use a particular method}. 
The first and second criteria are also in line with \textit{usability} and \textit{usefulness} as mentioned by~\citet{2022_Gericke_ElementsDesignMethod}.

\subsection{Evaluation against objectives}
As part of our design process, we have specified four objectives as outlined in Section~\ref{sec:researchApproach}. For the evaluation of our artefact, i.e. our evaluation method, we evaluate it against these objectives. In Table~\ref{tab:eval_obj}, we provide an overview of each objective and summarise how it is addressed within our evaluation method.

\small
\begin{table}[ht!]
\centering
  \caption{Overview of objectives defined in the design process and their achievement in our evaluation method.}
  \label{tab:eval_obj}
  \renewcommand{\arraystretch}{1.5}
 \begin{tabular}{p{0.3\textwidth}p{0.6\textwidth}}
    \toprule
    Objective & How is the objective achieved in our evaluation method?
    \\
    \midrule

    Integration of explicit and implicit validation.
      & Phase 2 of our evaluation method addresses explicit model validation.
      Phase 3 of our evaluation method addresses implicit model validation.
      \\

    Suitability for evaluation practices.
      & Our evaluation method is structured and step-based.
      \\

    Addressing both \glspl{IDM1}
      & Phase 2 of our evaluation method considers the selection of \glspl{QC} for both \glspl{IDM1} and their separated evaluation against these characteristics. 
      \\

    Adaptability to various model development stages.
      & Phase 3 of our evaluation method includes a modular step to adapt the testing to the current model development stage.
      \\
  \bottomrule
\end{tabular}
\end{table}
\normalsize

The first objective, integrating explicit and implicit validation, is achieved through targeted phases in our evaluation method. Phase 2 addresses explicit validation by evaluating the models against \glspl{QC}, ensuring \textit{conceptual model quality}. 
Phase 3 addresses implicit validation through system-level testing of model instances, ensuring \textit{system functionality}.

The second objective, ensuring suitability for evaluation practices, is achieved through a structured and step-based design. Unlike existing evaluation frameworks for conceptual models, which often focus on relating concepts to each other, our evaluation method connects every step and supports iterative evaluation cycles.

The third objective, which supports the evaluation of both \glspl{IDM1}, is achieved in Phase 2. Step 2.2 enables the selection of \glspl{QC} for both model types, while steps 2.3 i and 2.3 d address their separate evaluation against the selected characteristics.

The fourth objective, adaptability to various stages of development, is achieved in Phase 3. Through the consideration of formal and informal test types, our evaluation method addresses evaluation in both late and early stages of model development. Whereas formal test types enable evaluations for fully implemented test systems, informal test types, including simplified and conceptual testing, enable evaluations, where test systems are still under development.
The execution of these different test types is achieved by modular test step 3.2. This can be adapted to a method suitable for the selected test type or replaced by corresponding sub-steps. 

\subsection{Collection and evaluation of expert feedback}
Besides the evaluation of our evaluation method against defined objectives, we also collect expert feedback. Expert feedback is commonly used to evaluate methods~\citep{2022_Gericke_ElementsDesignMethod}. 
We provide an overview of the experts we consulted in Table~\ref{tab:eval_exp}. Since our evaluation method addresses a domain-specific problem, its evaluation requires expertise in the design and use of \glspl{IDM1}. 
Given that the availability of relevant experts was limited,
we combined purposive and convenience sampling~\citep{kelley2003good}. 
This resulted in the selection of nine qualified experts with sufficient domain knowledge to meaningfully evaluate our proposed method.

We collected the feedback through different means.
First, we conducted a focus group with three experts. We asked questions about the experts' experience in model use and design, and presented our evaluation method and consolidated set of \glspl{QC}.
Second, we collected written and oral feedback as part of a conference presentation. Two experts provided feedback in a short written report, while four experts participated in short semi-structured interviews. These interviews were guided, based on the three criteria for design method evaluation~\citep{2003_Moody_MethodEvaluationModel}: \textit{perceived ease of use}, \textit{perceived usefulness}, and \textit{intention to use}. In the following, we map the experts' feedback with these criteria. 

\small
\begin{table}[h!]
\centering
\caption{Overview of experts providing feedback on our evaluation method.}
\label{tab:eval_exp}
\renewcommand{\arraystretch}{1.5}

 \begin{tabular}{
     >{\centering\arraybackslash}p{0.02\textwidth}
     >{\centering\arraybackslash}p{0.26\textwidth}
     >{\centering\arraybackslash}p{0.22\textwidth}
     >{\centering\arraybackslash}p{0.22\textwidth}
     }
    \toprule
    ID & Background/Expertise & Experience in model use/ design & Feedback collection \\
    \midrule

    1 & Production Engineering & yes & Focus group \\

    2 & Software Engineering & yes & Focus group \\

     3 & Software Engineering & yes & Focus group \\

    4 & Energy Informatics  & yes & Written feedback \\

    5 & Energy Informatics  & yes & Written feedback \\

    6 & Computer Science & yes & Oral feedback \\
    
    7 & Electrical Engineering & yes & Oral feedback \\

    8 & Computer Science & yes & Oral feedback \\

    9 & Electrical Engineering and Computer Science & yes &  Oral feedback\\

  \bottomrule
\end{tabular}
\end{table}

\normalsize

All experts provided feedback on the \textit{perceived ease of use}.
In the focus group, feedback mainly addressed the clarity of the procedural steps and the articulation of our method's intended use. Written feedback highlighted the abstract nature of our evaluation method and its limited domain specificity. Expert 5 stated that \enquote{the domain-specifics of the proposed approach was not clear}. It was also noted that the example used was relatively simple, promoting suggestions to apply our evaluation method to a more complex scenario. Oral feedback reinforced this point by emphasising the need for concrete examples that clearly illustrate the method's applicability. Expert 9 stressed that examples of applicability would facilitate understanding of our proposed evaluation method. 
In response, we refined the description of our method. We expanded the supplementary information with more specific explanations on our method's core idea and its intended use. In particular, we demonstrate its domain-specific application through the procedure description and a smart grid-specific example application to models describing industrial energy flexibility. 
The feedback received underscores the need to view our method not merely as a visual representation (i.e., a flow chart), but as a composite artefact that includes descriptions of the procedure, the core idea, and also the articulation of the underlying implicit assumptions, as emphasised by~\citet{2022_Gericke_ElementsDesignMethod}.

The feedback we received about the \textit{perceived usefulness} of our evaluation method includes specific suggestions about its potential application. Expert 3 proposed a potential use in application lifecycle management processes within software development. The perceived usefulness is also expressed by Expert 5, who stated that our research is \enquote{interesting for the research community} and \enquote{up-to-date}. The results were described as \enquote{reasonable} and it was affirmed that \enquote{the work has value}.
In contrast, Expert 4 expressed concerns that our contribution remains at a high level.
In response, we clearly state the intended high-level nature of our method's representation to make it applicable to various \glspl{IDM1} used in different information exchange use cases.
This will allow model developers to combine and customise our method with specific test approaches, such as the CGMES Conformity Assessment Framework.
The overarching goal is to provide model developers with guidance in evaluating their designed models for various smart grid applications. 

The \textit{intention to use} our evaluation method was explicitly stated by five of the nine experts. Additionally, Expert 6 expressed a strong interest in our formalised evaluation method, noting the lack of similar approaches in the context of ontology development. This feedback confirms the practical relevance of our method and indicates a potential influence on ontology engineering approaches.

To summarise, nine experts with experience in the design and use of \glspl{IDM1} provided constructive feedback that led to refinements of our evaluation method's procedure descriptions, visual representation, and demonstration use cases. Notably, seven experts recognised our evaluation method's usefulness, of which five explicitly stated their intention to use it in potential future evaluations. This points to the practical relevance of our proposed evaluation method.

\section{Discussion} 
\label{sec:discussion}

Our proposed evaluation method for \glspl{IDM1} during their development process is discussed in this section. In particular, we examine its implications and limitations, and indicate potential directions for future investigation.

\subsection{Observed implications and lessons learned}

We observed several implications and derived lessons from the application of our evaluation method and from expert feedback.

First, we observed that after participating in the evaluation, model developers began to consider using \glspl{QC} during the subsequent model development process.
This change occurred after applying our method, as outlined in the example demonstrations in Section~\ref{subsec:procedure}. 
These observations indicate that our evaluation method might not only support model evaluation, but could also positively foster quality-conscious model development practices.

Second, we observed that \glspl{QC} derived from the reviewed literature tend to be more general in nature. Those identified through the observation-based approach appear to be more specific to \glspl{IDM1} used during the exchange of information.
New \glspl{IDM1} are expected to emerge in smart grid system interactions, for example, in the provision of \gls{DR}, and will require evaluation. 
We expect that these further evaluations will contribute to the evolution and refinement of the \gls{IDM1}-specific set of \glspl{QC}.
 
Third, in line with statements in the literature, we acknowledge that some of the \glspl{QC} in our consolidated set could be grouped together. \citet{1994_Lindland_UnderstandingQualityConceptual}, for example, suggests that \textit{consistency} and \textit{unambiguity} are part of \textit{validity and completeness}. However, our observations made during the application of our method align with findings in previous work. This indicates that specific \glspl{QC} are more useful when conducting practical evaluations than the broader, more general ones~\cite{1997_Shanks_QualityConceptualModelling}.

Fourth, although interdependencies between \glspl{QC} have already been discussed in work by~\citet{1995_Levitin_QualityDimensionsConceptual} and~\citet{1994_Moody_WhatMakesGood}, our observations relate to the newly proposed \gls{QC} \textit{singularity}, that is, one that has no redundant information. With the intention of increasing the \textit{understandability} of a model, we observed developers intentionally choosing to violate \textit{singularity}, thus allowing the representation of the model version in two different ways. Such decisions are domain- and use case-specific. These interdependencies also suggest that evaluation remains a creative process that cannot be fully automated. It often requires a considerable degree of expert intuition gained through extensive practical experience~\cite{1994_Moody_WhatMakesGood, 1994_Lindland_UnderstandingQualityConceptual}. 

Fifth, as outlined in Section~\ref{sec:eval}, experts from diverse fields (including computer science, electrical engineering, software engineering, and production engineering), evaluated the applicability of our evaluation method. They confirmed its relevance for use across these disciplines. Their feedback highlighted its potential to support both academic and industrial modelling activities. Furthermore, the step-wise procedure description of our method suggests its potential applicability to standardisation bodies involved in the specification and evaluation of \glspl{IDM1}.

Overall, although most of the literature on conceptual model evaluation was published in the 1990s and early 2000s, recent publications reflect renewed interest in this topic. This resurgence is driven by the emergence of novel application domains~\citep{2024_Taentzer_HowDefineQuality, 2024_Helskyaho_DefiningDataModel}. By providing structured guidance for practical evaluation processes, our method addresses recurring challenges in model development. Specifically, it addresses challenges that arise when new models emerge to describe new system interactions, such as the provision of \gls{DR} in smart grids. 

\subsection{Limitations and future work}

As is typical in design-oriented research, our evaluation method for \glspl{IDM1} in smart grid interactions is subject to several limitations. These include limited evidence of its generalisability, the qualitative nature of its evaluation, and a potential mismatch between the applied \glspl{QC} and the specific model context of information exchange. All of these could suggest potential future research directions.

First, while we demonstrated the effectiveness of our evaluation method through the evaluation of the \gls{EFIM} and \gls{EFDM}, its applicability to other \glspl{IDM1} has not yet been validated. Future research should apply our evaluation method to a broader range of \glspl{IDM1}. Doing so both within and beyond the smart grid domain would validate its generalisability. The healthcare sector represents a particularly promising application domain given its dependence on structured data exchange. 

Second, we conducted a qualitative evaluation. 
There is no quantifiable metric to determine the success of our evaluation method. 
This challenge is common to many design-oriented methods~\citep{2022_Gericke_ElementsDesignMethod}. Therefore, we evaluated our evaluation method against predefined objectives and analysed expert feedback using three evaluation criteria. Given our method's specific purpose and the expertise required for its evaluation, we relied on a small but well-qualified sample of domain experts.
Furthermore, the evaluation of \gls{EFIM} and \gls{EFDM} in the demonstration example was conducted at an intermediate development stage, where fully implemented test systems were not yet available. As a result, the evaluation relied on informal testing and did not allow for fully automated testing. Future research should examine \glspl{IDM1} in more advanced settings, where models integrated into test systems enable formal and automated testing.

Third, we applied a consolidated set of \glspl{QC} building on characteristics from the reviewed literature on conceptual model evaluation. Since these characteristics were originally developed for broader contexts such as database and software design, their relevance to \glspl{IDM1} in system interactions may be limited, and additional characteristics might be required. Not all \glspl{QC} (such as \textit{naturalness}) revealed defects in our example method demonstration. This could indicate the robustness of the model or suggest that certain characteristics are less useful for identifying issues in this context. Future studies should therefore evaluate additional \glspl{IDM1} within and beyond the smart grid domain to clarify the practical relevance of our proposed set of \glspl{QC}. This would determine whether further characteristics are needed.

\section{Conclusion}
\label{sec:conclusion}

The increasing digitalisation of the smart grid has led to the emergence of new \glspl{IDM1} to support automated information exchange. Evaluating these models during their development process is essential to ensure both \textit{conceptual model quality} and \textit{system functionality} before deployment. However, existing evaluation approaches either remain somewhat abstract or focused on post-deployment system testing. This leaves a gap in actionable guidance for model developers.

To address this gap, we designed a structured, three-phase evaluation method using \gls{DSR}. Our method combines explicit and implicit validation approaches and supports the evaluation of \glspl{IDM1} throughout their development. It includes model scope definition, conceptual evaluation based on quality characteristics, and system-level testing to assess functionality, specifically in terms of conformance and interoperability.

To support the explicit evaluation phase, we proposed a consolidated set of \glspl{QC}, integrating insights from the reviewed literature with three new characteristics derived through observation. These additions reflect practical concerns specific to information exchange in smart grid contexts.

Overall, our evaluation method provides a systematic and adaptable approach for assessing \glspl{IDM1} during development. It offers model developers a means to identify design flaws early, improve model reliability, and ensure alignment with system requirements.

\backmatter

\subsubsection*{Abbreviations}
\printglossary[type=\acronymtype, toctitle=Acronyms]

\section*{Declarations}

\subsubsection*{Availability of data and materials}
The expert feedback analysed in this study is not publicly available. Anonymised summaries may be provided upon reasonable request, internal check and compliance.

\subsubsection*{Competing interests}
The authors declare that they have no competing interests.

\subsubsection*{Funding}
This work has been supported by the Kopernikus-project “SynErgie” by the German Federal Ministry of Education and Research (BMBF) and by the Luxembourg National Research Fund (FNR) and PayPal, PEARL grant reference 13342933/Gilbert Fridgen. 

\subsubsection*{Author's contributions}
CvS: Conceptualisation, Investigation, Methodology, Data curation, Visualisation, Writing – original draft, Writing – review \& editing.
SPM: Supervision, Conceptualisation, Validation, Writing - review \& editing.
GF: Validation, Writing - review \& editing, Funding acquisition.
All authors read and approved the final manuscript.

\subsubsection*{Acknowledgements}
The authors gratefully would like to acknowledge the project supervision by the project management organisation Projektträger Jülich (PtJ) and the extensive discussions with the colleagues from the SIG EFDM of the SynErgie project. The authors would like to thank Dominik Vereno for his valuable feedback during the final revision. 

\subsubsection*{Declaration of Generative AI and AI-assisted technologies in the Writing Process}

During the preparation of this manuscript, the authors used Grammarly, ChatGPT, Copilot and DeepL to improve the clarity and readability of the text. After using these tools, the authors carefully reviewed and edited the content to ensure that the original meaning was preserved. The authors take full responsibility for the content of the publication.

\noindent

\bibliography{ref_bibtext}

@article{TUBALLA2016710,
  title   = {A review of the development of Smart Grid technologies},
  journal = {Renewable and Sustainable Energy Reviews},
  volume  = {59},
  pages   = {710-725},
  year    = {2016},
  issn    = {1364-0321},
  doi     = {10.1016/j.rser.2016.01.011},
  author  = {Maria Lorena Tuballa and Michael Lochinvar Abundo}
}

@inproceedings{7947600,
  author    = {Kuzlu, M. and Pipattanasompom, M. and Rahman, S.},
  booktitle = {2017 5th International Istanbul Smart Grid and Cities Congress and Fair (ICSG)},
  title     = {A comprehensive review of smart grid related standards and protocols},
  year      = {2017},
  pages     = {12-16},
  doi       = {10.1109/SGCF.2017.7947600}
}

@report{2015_openADRAlliance_OpenADR20Profile,
  title  = {OpenADR 2.0 Profile Specification B Profile},
  author = {{openADR Alliance}},
  date   = {2015},
  url = {https://www.openadr.org/specification},
  urldate = {08/05/2023}
}

@report{2023_EEBusInitiativee.V._EEBusSPINETechnical,
  title  = {EEBus SPINE Technical Report},
  author = {{EEBus Initiative e.V.}},
  date   = {2023},
  url = {https://www.eebus.org/specifications-media/},
  urldate = {19/01/2024}
}

@article{2019_Schott_GenericDataModel,
  title   = {A Generic Data Model for Describing Flexibility in Power Markets},
  author  = {Schott, Paul and Sedlmeir, Johannes and Strobel, Nina and Weber, Thomas and Fridgen, Gilbert and Abele, Eberhard},
  year    = {2019},
  journal = {Energies},
  volume  = {12},
  number  = {10},
  pages   = {1893},
  issn    = {1996-1073},
  doi     = {10.3390/en12101893},
  langid  = {english}
}

@article{2005_Moody_TheoreticalPracticalIssues,
  title      = {Theoretical and Practical Issues in Evaluating the Quality of Conceptual Models: Current State and Future Directions},
  shorttitle = {Theoretical and Practical Issues in Evaluating the Quality of Conceptual Models},
  author     = {Moody, Daniel L.},
  year       = {2005},
  journal    = {Data \& Knowledge Engineering},
  series     = {Quality in Conceptual Modeling},
  volume     = {55},
  number     = {3},
  pages      = {243--276},
  issn       = {0169-023X},
  doi        = {10.1016/j.datak.2004.12.005}
}

@article{2022_Cabot_ModelingShouldBe,
  title        = {Modeling Should Be an Independent Scientific Discipline},
  author       = {Cabot, Jordi and Vallecillo, Antonio},
  date         = {2022-12-01},
  journaltitle = {Software and Systems Modeling},
  shortjournal = {Softw Syst Model},
  volume       = {21},
  number       = {6},
  pages        = {2101--2107},
  publisher    = {Springer Berlin Heidelberg},
  issn         = {1619-1374},
  doi          = {10.1007/s10270-022-01035-8},
  issue        = {6},
  langid       = {english}
}

@report{2014_CEN_SGAMUserManual,
  title  = {SGAM User Manual - Applying, Testing \& Refining the Smart Grid Architecture Model (SGAM), Version 3.0},
  author = {{CEN} and {CENELEC} and {ETSI}},
  year   = {2014},
  url    = {https://syc-se.iec.ch/wp-content/uploads/2019/10/SGCG_Methodology_SGAMUserManual.pdf},
  urldate = {31/07/2024}
}

@article{1995_Levitin_QualityDimensionsConceptual,
  title   = {Quality Dimensions of a Conceptual View},
  author  = {Levitin, Anany and Redman, Thomas},
  year    = {1995},
  journal = {Information Processing \& Management},
  volume  = {31},
  number  = {1},
  pages   = {81--88},
  issn    = {0306-4573},
  doi     = {10.1016/0306-4573(95)80008-H}
}

@article{1978_Tsichritzis_ANSIX3SPARC,
  title        = {The ANSI/X3/SPARC DBMS Framework Report of the Study Group on Database Management Systems},
  author       = {Tsichritzis, Dennis and Klug, Anthony},
  date         = {1978-01-01},
  journaltitle = {Information Systems},
  shortjournal = {Information Systems},
  volume       = {3},
  number       = {3},
  pages        = {173--191},
  issn         = {0306-4379},
  doi          = {10.1016/0306-4379(78)90001-7}
}

@article{1994_Lindland_UnderstandingQualityConceptual,
  title   = {Understanding Quality in Conceptual Modeling},
  author  = {Lindland, O.I. and Sindre, G. and Solvberg, A.},
  year    = {1994},
  journal = {IEEE Software},
  volume  = {11},
  number  = {2},
  pages   = {42--49},
  issn    = {1937-4194},
  doi     = {10.1109/52.268955}
}

@book{2012_Krogstie_ModelbasedDevelopmentEvolution,
  title      = {Model-Based Development and Evolution of Information Systems: A Quality Approach},
  shorttitle = {Model-Based Development and Evolution of Information Systems},
  author     = {Krogstie, John},
  year       = {2012},
  publisher  = {Springer},
  address    = {New York},
  isbn       = {978-1-4471-2935-6},
  doi        = {10.1007/978-1-4471-2936-3},
  langid     = {english}
}

@article{1999_Lee_InformationModelingDesign,
  title      = {Information Modeling: From Design to Implementation},
  shorttitle = {Information Modeling},
  author     = {Lee, Yung-Tsun T.},
  year       = {1999},
  month      = sep,
  journal    = {IEEE Transactions on Robotics and Automation},
  publisher  = {Yung-Tsun T. Lee},
  urldate    = {2023/06/11},
  langid     = {english},
  url        = {https://www.nist.gov/publications/information-modeling-design-implementation}
}

@techreport{2003_Pras_DifferenceInformationModels,
  title       = {On the Difference between Information Models and Data Models},
  author      = {Pras, A. and Schoenwaelder, J.},
  year        = {2003},
  number      = {RFC3444},
  institution = {RFC Editor},
  doi         = {10.17487/rfc3444}
}

@report{2014_ObjectManagementGroup_ModelDrivenArchitecture,
  title  = {Model Driven Architecture (MDA) MDA Guide Rev. 2.0},
  author = {{Object Management Group}},
  date   = {2014},
  url    = {https://www.omg.org/cgi-bin/doc?ormsc/14-06-01},
  urldate = {2025/03/28}
}

@article{2003_Moody_ImprovingQualityData,
  title      = {Improving the Quality of Data Models: Empirical Validation of a Quality Management Framework},
  shorttitle = {Improving the Quality of Data Models},
  author     = {Moody, Daniel L and Shanks, Graeme G},
  year       = {2003},
  journal    = {Information Systems},
  volume     = {28},
  number     = {6},
  pages      = {619--650},
  issn       = {0306-4379},
  doi        = {10.1016/S0306-4379(02)00043-1}
}

@article{2019_Uslar_ApplyingSmartGrid,
  title        = {Applying the Smart Grid Architecture Model for Designing and Validating System-of-Systems in the Power and Energy Domain: A European Perspective},
  shorttitle   = {Applying the Smart Grid Architecture Model for Designing and Validating System-of-Systems in the Power and Energy Domain},
  author       = {Uslar, Mathias and Rohjans, Sebastian and Neureiter, Christian and Pröstl Andrén, Filip and Velasquez, Jorge and Steinbrink, Cornelius and Efthymiou, Venizelos and Migliavacca, Gianluigi and Horsmanheimo, Seppo and Brunner, Helfried and Strasser, Thomas I.},
  date         = {2019-01-15},
  journaltitle = {Energies},
  shortjournal = {Energies},
  volume       = {12},
  number       = {2},
  pages        = {258},
  issn         = {1996-1073},
  doi          = {10.3390/en12020258},
  langid       = {english}
}

@book{2020_Stuckenholz_BasiswissenEnergieinformatikLehr,
  title      = {{Basiswissen} {Energieinformatik}: {Ein} {Lehr-} {und} {Arbeitsbuch} {für} {Studierende} {und} {Anwender} [{Basic} {knowledge} {of} {energy} {informatics}: {A} {text-} {and} {workbook} {for} {students} {and} {users}]},
  shorttitle = {{Basiswissen Energieinformatik}},
  author     = {Stuckenholz, Alexander},
  year       = {2020},
  publisher  = {Springer Fachmedien Wiesbaden},
  address = {Wiesbaden}, 
  doi        = {10.1007/978-3-658-31809-3},
  langid     = {ngerman}
}

@article{1996_Rahimifard_MethodologyDevelopEXPRESS,
  title        = {A Methodology to Develop EXPRESS Data Models},
  author       = {Rahimifard, S. and Newman, S.T.},
  year         = {1996},
  volume       = {9},
  doi          = {10.1080/095119296131814},
  langid       = {english},
  journaltitle = {International Journal of Computer Integrated Manufacturing}
}

@standard{2017_ISO/IEC/IEEE24765:2017E_ISOIECIEEE,
  title   = {ISO/IEC/IEEE 24765:2017(E) International Standard - Systems and software engineering–Vocabulary},
  author  = {{ISO/IEC/IEEE}},
  doi     = {10.1109/IEEESTD.2017.8016712},
  year    = {2017},
  langid  = {english}
}

@article{2005_Torraco_WritingIntegrativeLiterature,
  title      = {Writing Integrative Literature Reviews: Guidelines and Examples},
  shorttitle = {Writing Integrative Literature Reviews},
  author     = {Torraco, Richard J.},
  date       = {2005-09-01},
  journaltitle = {Human Resource Development Review},
  volume     = {4},
  number     = {3},
  pages      = {356--367},
  publisher  = {SAGE Publications},
  issn       = {1534-4843},
  doi        = {10.1177/1534484305278283}
}

@article{2006_Green_WritingNarrativeLiterature,
  title      = {Writing Narrative Literature Reviews for Peer-Reviewed Journals: Secrets of the Trade},
  shorttitle = {Writing Narrative Literature Reviews for Peer-Reviewed Journals},
  author     = {Green, Bart N. and Johnson, Claire D. and Adams, Alan},
  date       = {2006-09-01},
  journaltitle = {Journal of Chiropractic Medicine},
  shortjournal = {Journal of Chiropractic Medicine},
  volume     = {5},
  number     = {3},
  pages      = {101--117},
  issn       = {1556-3707},
  doi        = {10.1016/S0899-3467(07)60142-6}
}

@article{2023_Cronin_WhyHowIntegrative,
  title        = {The Why and How of the Integrative Review},
  author       = {Cronin, Matthew A. and George, Elizabeth},
  date         = {2023-01-01},
  journaltitle = {Organizational Research Methods},
  volume       = {26},
  number       = {1},
  pages        = {168--192},
  publisher    = {SAGE Publications Inc},
  issn         = {1094-4281},
  doi          = {10.1177/1094428120935507}
}

@article{SNYDER2019333,
  title   = {Literature review as a research methodology: An overview and guidelines},
  journal = {Journal of Business Research},
  volume  = {104},
  pages   = {333-339},
  year    = {2019},
  issn    = {0148-2963},
  doi     = {10.1016/j.jbusres.2019.07.039},
  author  = {Hannah Snyder}
}

@article{grant2009typology,
  title     = {A typology of reviews: an analysis of 14 review types and associated methodologies},
  author    = {Grant, Maria J and Booth, Andrew},
  journal   = {Health information \& libraries journal},
  volume    = {26},
  number    = {2},
  pages     = {91--108},
  year      = {2009},
  publisher = {Wiley Online Library},
  doi       = {10.1111/j.1471-1842.2009.00848.x}
}

@online{_EuropeanCommission_CORDISEUResearch,
  title   = {CORDIS - EU Research Results},
  author  = {{European Commission}},
  url     = {https://cordis.europa.eu},
  urldate = {2025/05/14},
  date    = {2025},
  langid  = {english}
}

@article{2014_Callahan_WritingLiteratureReviews,
  title        = {Writing Literature Reviews: A Reprise and Update},
  shorttitle   = {Writing Literature Reviews},
  author       = {Callahan, Jamie L.},
  date         = {2014},
  journaltitle = {Human Resource Development Review},
  volume       = {13},
  number       = {3},
  pages        = {271--275},
  publisher    = {SAGE Publications},
  issn         = {1534-4843},
  doi          = {10.1177/1534484314536705},
  langid       = {english}
}

@online{2025_entsoe_CommonInformationModel,
  title   = {Common Information Model (CIM) for Energy Markets},
  author  = {{ENTSO-E}},
  date    = {2025},
  url     = {https://www.entsoe.eu/digital/common-information-model/cim-for-energy-markets/},
  urldate = {2025/04/20}
}

@online{2019_openADRAlliance_OpenADR20bReceives,
  title        = {OpenADR 2.0b Receives Approval as IEC Standard},
  author       = {{openADR Alliance}},
  date         = {2019},
  url          = {https://www.openadr.org/index.php?option=com_content&view=article&id=174:openadr-2-0b-receives-approval-as-iec-standard&Itemid=121},
  urldate      = {2025/04/20},
  organization = {OpenADR 2.0b Specification Receives Full Approval as International Electrotechnical Commission (IEC) Standard}
}

@inproceedings{2009_Mehmood_DataQualityConceptual,
  author    = {Mehmood, Kashif and Cherfi, Samira and Wattiau, Isabelle},
  title     = {Data Quality through Conceptual Model Quality: Reconciling Researchers and Practitioners through a Customizable Quality Model},
  booktitle = {Proceedings of the 2009 International Conference on Information Quality (ICIQ 2009)},
  year      = {2009},
  pages     = {61--74},
  url       = {https://hal.science/hal-01125702},
  urldate = {2024/05/17}
}

@article{2009_Batini_MethodologiesDataQuality,
  title        = {Methodologies for Data Quality Assessment and Improvement},
  author       = {Batini, Carlo and Cappiello, Cinzia and Francalanci, Chiara and Maurino, Andrea},
  date         = {2009-07},
  journaltitle = {ACM Computing Surveys},
  shortjournal = {ACM Comput. Surv.},
  volume       = {41},
  number       = {3},
  pages        = {1--52},
  issn         = {0360-0300, 1557-7341},
  doi          = {10.1145/1541880.1541883}
}

@article{2024_Helskyaho_DefiningDataModel,
  title        = {Defining Data Model Quality Metrics for Data Vault 2.0 Model Evaluation},
  author       = {Helskyaho, Heli and Ruotsalainen, Laura and Männistö, Tomi},
  date         = {2024-02},
  journaltitle = {Inventions},
  volume       = {9},
  number       = {1},
  pages        = {21},
  publisher    = {Multidisciplinary Digital Publishing Institute},
  issn         = {2411-5134},
  doi          = {10.3390/inventions9010021},
  issue        = {1}
}

@inproceedings{2013_Krogstie_QualityConceptualData,
  title     = {Quality of Conceptual Data Models},
  booktitle = {International Conference on Informatics and Semiotics in Organisations},
  author    = {Krogstie, John},
  year      = {2013},
  langid    = {english},
  url       = {https://folk.idi.ntnu.no/krogstie/publications/2013/ICISO/iciso-2013-art.pdf},
  urldate   = {2024-02-06}
}

@article{1990_Wand_OntologicalModelInformation,
  title        = {An Ontological Model of an Information System},
  author       = {Wand, Y. and Weber, R.},
  date         = {1990-11},
  journaltitle = {IEEE Transactions on Software Engineering},
  volume       = {16},
  number       = {11},
  pages        = {1282--1292},
  issn         = {1939-3520},
  doi          = {10.1109/32.60316}
}

@article{2012_Nelson_ConceptualModelingQuality,
  title        = {A Conceptual Modeling Quality Framework},
  author       = {Nelson, H. James and Poels, Geert and Genero, Marcela and Piattini, Mario},
  date         = {2012-03-01},
  journaltitle = {Software Quality Journal},
  shortjournal = {Software Qual J},
  volume       = {20},
  number       = {1},
  pages        = {201--228},
  issn         = {1573-1367},
  doi          = {10.1007/s11219-011-9136-9},
  langid       = {english}
}

@article{2009_Mohagheghi_DefinitionsApproachesModel,
  title      = {Definitions and Approaches to Model Quality in Model-Based Software Development -- A Review of Literature},
  author     = {Mohagheghi, Parastoo and Dehlen, Vegard and Neple, Tor},
  year       = {2009},
  journal    = {Information and Software Technology},
  series     = {Quality of UML Models},
  volume     = {51},
  number     = {12},
  pages      = {1646--1669},
  issn       = {0950-5849},
  doi        = {10.1016/j.infsof.2009.04.004}
}

@article{2021_Haug_UnderstandingDifferencesData,
  title      = {Understanding the Differences across Data Quality Classifications: A Literature Review and Guidelines for Future Research},
  shorttitle = {Understanding the Differences across Data Quality Classifications},
  author     = {Haug, Anders},
  year       = {2021},
  journal    = {Industrial Management \& Data Systems},
  volume     = {121},
  number     = {12},
  pages      = {2651--2671},
  publisher  = {Emerald Publishing Limited},
  issn       = {0263-5577},
  doi        = {10.1108/IMDS-12-2020-0756}
}

@inproceedings{2003_Moody_MeasuringQualityData,
  title     = {Measuring the Quality of Data Models: An Empirical Evaluation of the Use of Quality Metrics in Practice},
  author    = {Moody, Daniel L},
  year      = {2003},
  langid    = {english},
  url       = {https://aisel.aisnet.org/ecis2003/78},
  urldate   = {2024/02/06},
  booktitle = {ECIS 2003 Proceedings}
}

@inproceedings{1997_Shanks_QualityConceptualModelling,
  author    = {Shanks, Graeme and Darke, Peta},
  title     = {Quality in Conceptual Modelling: Linking Theory and Practice},
  booktitle = {Proceedings of the Pacific Asia Conference on Information Systems (PACIS 1997)},
  year      = {1997},
  url       = {http://aisel.aisnet.org/pacis1997},
  urldate   = {2024/02/06}
}

@article{2024_Taentzer_HowDefineQuality,
  title      = {{How to Define the Quality of Data and Software Models? A Data Quality Perspective}},
  shorttitle = {How to Define the Quality of Data and Software Models?},
  author     = {Taentzer, Gabriele and Kesper, Arno and Matoni, Markus},
  year       = {2024},
  publisher  = {Gesellschaft für Informatik e.V.},
  doi        = {10.18420/MODELLIERUNG2024-WS-010},
  langid     = {english}
}

@inproceedings{1994_Moody_WhatMakesGood,
  author    = {Moody, Daniel L. and Shanks, Graeme G.},
  title     = {What Makes a Good Data Model? Evaluating the Quality of Entity Relationship Models},
  booktitle = {Entity--Relationship Approach --- ER'94: Business Modelling and Re-Engineering},
  editor    = {Loucopoulos, Pericles},
  year      = {1994},
  series    = {Lecture Notes in Computer Science},
  pages     = {94--111},
  publisher = {Springer},
  address   = {Berlin, Heidelberg},
  doi       = {10.1007/3-540-58786-1_75}
}

@inproceedings{1995_Krogstie_DeeperUnderstandingQuality,
  title     = {Towards a Deeper Understanding of Quality in Requirements Engineering},
  booktitle = {Advanced Information Systems Engineering},
  author    = {Krogstie, John and Lindland, Odd Ivar and Sindre, Guttorm},
  editor    = {Iivari, Juhani and Lyytinen, Kalle and Rossi, Matti},
  year      = {1995},
  pages     = {82--95},
  publisher = {Springer},
  address   = {Berlin, Heidelberg},
  doi       = {10.1007/3-540-59498-1_239},
  isbn      = {978-3-540-49290-0},
  langid    = {english}
}

@report{2018_Papaioannou_SmartGridInteroperability,
  title       = {Smart Grid Interoperability Testing Methodology},
  author      = {Papaioannou, I. and Tarantola, S. and Lucas, A. and Kotsakis, E. and Marinopoulos, A. and Olariaga Guardiola, M. and Masera, M.},
  date        = {2018},
  institution = {Publications Office of the European Union},
  location    = {LU},
  url         = {https://data.europa.eu/doi/10.2760/08049},
  urldate     = {21/09/2023}
}

@online{2016_NIST_ConformanceTesting,
  title        = {Conformance Testing},
  author       = {{NIST}},
  date         = {2016},
  url          = {https://www.nist.gov/itl/ssd/information-systems-group/conformance-testing},
  urldate      = {2025/01/07},
  langid       = {english},
  organization = {NIST}
}

@report{2021_Gopstein_NISTFrameworkRoadmapa,
  title       = {NIST Framework and Roadmap for Smart Grid Interoperability Standards, Release 4.0},
  author      = {Gopstein, Avi and Nguyen, Cuong and O'Fallon, Cheyney and Hastings, Nelson and Wollman, David},
  year        = {2021},
  number      = {NIST SP 1108r4},
  institution = {National Institute of Standards and Technology},
  doi         = {10.6028/NIST.SP.1108r4}
}

@article{2021_Fernandez-Izquierdo_ConformanceTestingOntologies,
  title        = {Conformance Testing of Ontologies through Ontology Requirements},
  author       = {Fernández-Izquierdo, Alba and García-Castro, Raúl},
  date         = {2021},
  journaltitle = {Engineering Applications of Artificial Intelligence},
  shortjournal = {Engineering Applications of Artificial Intelligence},
  volume       = {97},
  pages        = {104026},
  issn         = {09521976},
  doi          = {10.1016/j.engappai.2020.104026}
}

@standard{_ISO/IEC_ISOIEC96461,
  title      = {ISO/IEC 9646-1:1994 Information technology — Open Systems Interconnection — Conformance testing methodology and framework},
  shorttitle = {ISO/IEC 9646-1},
  author     = {{ISO/IEC}},
  url        = {https://www.iso.org/standard/17473.html},
  urldate    = {2025/01/12},
  langid     = {english},
  year       = {1994}
}

@online{_ETSI_ConformanceTesting,
  title        = {Conformance Testing},
  author       = {{ETSI}},
  date         = {2024},
  url          = {https://portal.etsi.org/Services/Centre-for-Testing-Interoperability/ETSI-Approach/Conformance},
  urldate      = {2025/01/08},
  organization = {Testing, Interoperability and Technical Quality}
}

@article{2003_Grabowski_IntroductionTestingTest,
  title        = {An Introduction to the Testing and Test Control Notation (TTCN-3)},
  author       = {Grabowski, Jens and Hogrefe, Dieter and Réthy, György and Schieferdecker, Ina and Wiles, Anthony and Willcock, Colin},
  date         = {2003-06},
  journaltitle = {Computer Networks},
  shortjournal = {Computer Networks},
  volume       = {42},
  number       = {3},
  pages        = {375--403},
  issn         = {13891286},
  doi          = {10.1016/S1389-1286(03)00249-4}
}

@report{2024_Kung_EvolutionInteroperabilityStandards,
  title  = {Evolution of Interoperability Standards},
  author = {Kung, Antonio and Simmon, Eric and Lambert, Eric and Tobich, Karim and Alve, Jukka and Lahrin, Tobbe and Spanakis, Emmanouil and An, Xiaomi and Board, Dave and Coallier, François and Mulquin, Michael and Castro, Raul},
  date   = {2024},
  url    = {https://aioti.eu/wp-content/uploads/IntNet-Stand.ICT-AIOTI-Report-on-the-evolution-of-interoperability-standards.pdf},
  urldate = {2025/03/31}
}

@inproceedings{2025_Strasser_InteroperabilityTestingSmart,
  title     = {Towards Interoperability Testing of Smart Energy Systems – an Overview and Discussion of Possibilities},
  booktitle = {IET Conference Proceedings},
  author    = {Strasser, Thomas I. and Widl, Edmund and Kuchenbuch, René A. and Lázaro-Elorriaga, Laura and Laraudogoitia, Borja Tellado and Ginocchi, Mirko and Penthong, Thanakorn and Ponci, Ferdinanda and Gyrard, Amelie and Kung, Antonio and Mac Gregor, Carlos A. and Montero, Carmen Garcia and Algaba, Eduardo Relano},
  date      = {2025-02},
  volume    = {2024},
  pages     = {263--268},
  publisher = {The Institution of Engineering and Technology},
  doi       = {10.1049/icp.2024.4670}
}

@report{2017_entsoe_CGMESConformityAssessment,
  title  = {CGMES Conformity Assessment Framework},
  author = {{ENTSO-E}},
  date   = {2017},
  url    = {https://eepublicdownloads.entsoe.eu/clean-documents/CIM_documents/Grid_Model_CIM/171016%20ENTSO-E%20ConformityAssessmentFramework%20V2.pdf},
  urldate = {2025/04/27}
}

@online{_EuropeanCommissionsDIGIT_InteroperabilityTestBed,
  type         = {GitHub},
  title        = {Interoperability Test Bed},
  author       = {{European Commission’s DIGIT}},
  url          = {https://github.com/ISAITB},
  urldate      = {2025/01/12},
  date         = {2025},
  langid       = {english},
  organization = {European Commission’s DIGIT}
}

@online{_JRCSmartElectricitySystems_SmartGridDesign,
  title   = {Smart Grid Design of Interoperability Tests (SG-DoIT)},
  author  = {{JRC Smart Electricity Systems}},
  url     = {https://ses.jrc.ec.europa.eu/sgdoit},
  urldate = {2025/04/19}
}

@report{2022_Sauermann_WHITEPAPERSECTOR,
  title  = {WHITE PAPER FOR A SECTOR NEUTRAL INTEROPERABILITY PROCESS},
  author = {Sauermann, Stefan and Birtanov, Adletkhan and Urbauer, Philipp and Rohatsch, Lukas and Frohner, Matthias and Wagner, Adrian and Anderluh, Alexandra and Michelberger, Frank},
  date   = {2022},
  url    = {https://phaidra.fhstp.ac.at/o:5193},
  urldate = {2025/01/03}
}

@online{2023_IHE-Europe_GazelleIHEEurope,
  title   = {Gazelle | IHE Europe},
  author  = {{IHE-Europe}},
  date    = {2023},
  url     = {https://www.ihe-europe.net/testing-IHE/gazelle},
  urldate = {2025/05/12}
}

@article{2019_Heussen_ERIGridHolisticTest,
  title        = {ERIGrid Holistic Test Description for Validating Cyber-Physical Energy Systems},
  author       = {Heussen, Kai and Steinbrink, Cornelius and Abdulhadi, Ibrahim F. and Nguyen, Van Hoa and Degefa, Merkebu Z. and Merino, Julia and Jensen, Tue V. and Guo, Hao and Gehrke, Oliver and Bondy, Daniel Esteban Morales and Babazadeh, Davood and Pröstl Andrén, Filip and Strasser, Thomas I.},
  date         = {2019-01},
  journaltitle = {Energies},
  volume       = {12},
  number       = {14},
  pages        = {2722},
  publisher    = {Multidisciplinary Digital Publishing Institute},
  issn         = {1996-1073},
  doi          = {10.3390/en12142722}
}

@article{2007_Peffers_DesignScienceResearchc,
  title        = {A Design Science Research Methodology for Information Systems Research},
  author       = {Peffers, Ken and Tuunanen, Tuure and Rothenberger, Marcus A. and Chatterjee, Samir},
  year         = {2007},
  journal      = {Journal of Management Information Systems},
  volume       = {24},
  number       = {3},
  eprint       = {40398896},
  eprinttype   = {jstor},
  pages        = {45--77},
  langid       = {english},
  url          = {http://www.jstor.org/stable/40398896}
}

@thesis{2023_PotencianoMenci_SYSTEMORGANIZATIONOPERATION,
  title       = {System organization and operation in the context of local flexibility markets at distribution level},
  author      = {Potenciano Menci, Sergio},
  date        = {2023},
  institution = {University of Luxembourg},
  type        = {Ph.D. Thesis},
  url = {https://orbilu.uni.lu/bitstream/10993/60001/1/PhD_Thesis_SPM_orbilu.pdf},
  urldate = {2024/11/21}
}

@standard{2024_IEC_IEC62559Use,
  title   = {IEC 62559 – Use Case Methodology},
  author  = {{IEC}},
  date    = {2024},
  url     = {https://syc-se.iec.ch/deliveries/iec-62559-use-cases/},
  urldate = {2024/09/09}
}

@software{2023_Lindner_EnergyFlexibilityDataa,
  title         = {Energy Flexibility Data Model (EFDM)},
  author        = {Lindner, Martin and Koch, Tobias},
  date          = {2023-11-01},
  doi           = {10.5281/zenodo.8409627},
  organization  = {Zenodo}
}

@article{2025_vanStiphoudt_EnergySynchronizationPlatform,
  title        = {The Energy Synchronization Platform Concept in the Model Region Augsburg to Enable and Streamline Automated Industrial Demand Response},
  author       = {van Stiphoudt, Christine and Potenciano Menci, Sergio and Kaymakci, Can and Wenninger, Simon and Bauer, Dennis and Duda, Sebastian and Fridgen, Gilbert and Sauer, Alexander},
  date         = {2025-06},
  journaltitle = {Applied Energy},
  shortjournal = {Applied Energy},
  volume       = {388},
  pages        = {125455},
  issn         = {03062619},
  doi          = {10.1016/j.apenergy.2025.125455},
  langid       = {english}
}

@report{2012_ETSI_MethodsTestingSpecification,
  title    = {Methods for Testing and Specification (MTS); Standards Engineering Process; A Handbook of Validation Methods},
  author   = {{ETSI}},
  date     = {2012},
  number   = {EG 201 015 V2.1.1},
  url      = {https://www.etsi.org/deliver/etsi_eg/201000_201099/201015/02.01.01_60/eg_201015v020101p.pdf},
  langid   = {english},
  urldate = {08/04/2025}
}

@report{1999_ETSI_MethodsTestingSpecification,
  title  = {Methods for Testing and Specification (MTS); Planning for Validation and Testing in the Standards-Making Process},
  author = {{ETSI}},
  date   = {1999},
  number = {ETSI EG 202 107},
  url    = {https://www.etsi.org/deliver/etsi_eg/202100_202199/202107/01.01.01_60/eg_202107v010101p.pdf},
  urldate = {2025/04/08}
}

@article{2022_Gericke_ElementsDesignMethod,
  title        = {Elements of a Design Method – a Basis for Describing and Evaluating Design Methods},
  author       = {Gericke, Kilian and Eckert, Claudia and Stacey, Martin},
  date         = {2022-01},
  journaltitle = {Design Science},
  volume       = {8},
  pages        = {e29},
  issn         = {2053-4701},
  doi          = {10.1017/dsj.2022.23},
  langid       = {english}
}

@inproceedings{2003_Moody_MethodEvaluationModel,
  title     = {The Method Evaluation Model: A Theoretical Model for Validating Information Systems Design Methods},
  author    = {Moody, Daniel L},
  year      = {2003},
  langid    = {english},
  url       = {https://aisel.aisnet.org/ecis2003/79},
  urldate   = {2024/02/06},
  booktitle = {ECIS 2003 Proceedings}
}

@inproceedings{2026_vanStiphoudt_ExAnteEvaluationApproaches,
  title     = {Ex-Ante Evaluation Approaches Within the Design Process of Information and Data Models},
  booktitle = {Business Modeling and Software Design},
  author    = {{van Stiphoudt}, Christine and Potenciano Menci, Sergio and Fridgen, Gilbert},
  year      = {2026},
  pages     = {220--229},
  publisher = {Springer},
  address = {Cham},
  issn      = {1865-1356},
  doi       = {10.1007/978-3-031-98033-6_16},
  urldate   = {2025-07-14},
  isbn      = {978-3-031-98033-6},
  langid    = {english}
}

@incollection{2012_Watson_EnergyInformaticsInitial,
  title      = {Energy Informatics: Initial Thoughts on Data and Process Management},
  shorttitle = {Energy Informatics},
  booktitle  = {Green Business Process Management},
  author     = {Watson, Richard T. and Howells, Jeffrey and Boudreau, Marie-Claude},
  editor     = {{vom Brocke}, Jan and Seidel, Stefan and Recker, Jan},
  year       = {2012},
  pages      = {147--159},
  publisher  = {Springer Berlin Heidelberg},
  address    = {Berlin, Heidelberg},
  doi        = {10.1007/978-3-642-27488-6_9},
  isbn       = {978-3-642-27487-9 978-3-642-27488-6},
  langid     = {english}
}

@inproceedings{2015_USLAR_EnergyInformaticsDefinition,
  title      = {Energy Informatics: Definition, State-of-the-Art and New Horizons},
  shorttitle = {Energy Informatics},
  booktitle  = {ComForEn 2015},
  author     = {USLAR, {MATHIAS}},
  year       = {2015},
  volume     = {5},
  address    = {Wien},
  urldate    = {2025-06-02},
  langid     = {english},
  url = {{https://www.researchgate.net/publication/281006115}}
}

@online{SynErgie,
  title   = {Projektziel [{project} {goal}]},
  author  = {{SynErgie}},
  url     = {https://synergie-projekt.de/ueber-synergie/projektziel},
  urldate = {2025/10/10},
  date    = {2025},
  langid  = {german}
}

@article{kelley2003good,
  title={Good practice in the conduct and reporting of survey research},
  author={Kelley, Kate and Clark, Belinda and Brown, Vivienne and Sitzia, John},
  journal={International Journal for Quality in health care},
  volume={15},
  number={3},
  pages={261--266},
  year={2003},
  publisher={Oxford University Press}, 
  doi = {https://doi.org/10.1093/intqhc/mzg031}
}

@online{GitLab_EFDM,
  title   = {Energy Flexibility Data Models - Tickets},
  author  = {{SynErgie}},
  url     = {https://git.ptw.maschinenbau.tu-darmstadt.de/eta-fabrik/public/energy_flexibility_data_model/-/issues},
  urldate = {2025/12/10},
  date    = {2025},
  langid  = {german}
}

\end{document}